\documentclass[journal]{IEEEtran}

\usepackage{etoolbox}
\makeatletter
\patchcmd{\@makecaption}
{\scshape}
{}
{}
{}
\makeatletter
\patchcmd{\@makecaption}
{\\}
{.\ }
{}
{}
\makeatother
\usepackage{diagbox}
\usepackage{soul} 
\usepackage{ragged2e} 
\usepackage{resizegather}
\usepackage{subcaption}
\usepackage{multirow}
\usepackage{multicol}
\usepackage[noadjust]{cite}  
\usepackage[table,xcdraw]{xcolor}
\usepackage{tikz,tkz-tab} 
\usepackage[ruled,vlined]{algorithm2e}
\usepackage{float}
\usepackage[normalem]{ulem} 
\usepackage{algcompatible}
\usepackage{cuted}
\usepackage{xpatch}
\usepackage[author={Antoine},color=yellow]{pdfcomment}
\usepackage[utf8]{inputenc}
\usepackage[english]{babel}
\usepackage{amsmath} 
\usepackage{amsthm} 
\usepackage[cmintegrals]{newtxmath}
\usepackage[utf8]{inputenc}
\usepackage{graphicx}
\usepackage{svg}
\graphicspath{{./Figures/}{./ResultFigs/}} 
\usepackage{textcomp} 
\usepackage{mathtools}
\usepackage{cuted}
\usepackage{cases}
\usepackage{caption}
\usepackage{tabularx} 

\usepackage{multirow}
\DeclareMathOperator{\sign}{sgn}
\usepackage{cleveref}
\usepackage{comment}
\usepackage{balance}
\newcommand{\commentaire}[1]{}

\definecolor{lred}{RGB}{233,150,122}
\definecolor{mred}{RGB}{205,92,92}
\definecolor{dred}{RGB}{128,0,0}

\newtheorem{proposition}{Proposition}

\usepackage{xparse} \DeclarePairedDelimiterX{\Iintv}[1]{\llbracket}{\rrbracket}{\iintvargs{#1}}
\NewDocumentCommand{\iintvargs}{>{\SplitArgument{1}{,}}m}
{\iintvargsaux#1} %
\NewDocumentCommand{\iintvargsaux}{mm} {#1\mkern1.5mu..\mkern1.5mu#2}
\newcounter{loop@idx}  
\def\qhops#1{%
	\setcounter{loop@idx}{1}%
	\loop
	\quad%
	\ifnum\value{loop@idx}<#1
	\addtocounter{loop@idx}{1}
	\repeat
}
\makeatletter
\xpatchcmd{\algorithmic}
{\ALG@tlm\z@}{\leftmargin\z@\ALG@tlm\z@}
{}{}
\makeatother
\newcommand{\revised}[1]{\textcolor{magenta}{#1}}
\begin{document}
	\title{Optimal Resource Allocation for Full-Duplex IoT Systems Underlaying Cellular Networks with Mutual SIC NOMA}

	\author{~Antoine Kilzi,~Joumana~Farah,~Charbel~Abdel~Nour,~Catherine~Douillard
		\thanks{J. Farah is with the Department of Electricity and Electronics,
			Faculty of Engineering, Lebanese University, Roumieh, Lebanon
			(joumanafarah@ul.edu.lb).}
		\thanks{A. Kilzi, C. Abdel Nour and C. Douillard are with Institut Mines-Telecom, CNRS
			UMR 6285 Lab-STICC, 29238 Brest, France, (email: antoine.kilzi@imt-atlantique.fr; charbel.abdelnour@imt-atlantique.fr; catherine.douillard@imt-atlantique.fr).}
		\thanks{This work has been funded with support from IMT Atlantique and the Lebanese University Research Support Program.}}
	\maketitle

	\begin{abstract}
		Device-to-device (D2D) and non-orthogonal multiple access (NOMA) are promising technologies to meet the challenges of the next generations of mobile communications in terms of network density and diversity for internet of things (IoT) services.  This paper tackles the problem of maximizing the D2D sum-throughput in an IoT system underlaying a cellular network, through optimal channel and power allocation. NOMA is used to manage the interference between cellular users and full-duplex (FD) IoT devices. To this aim, mutual successive interference cancellation (SIC) conditions are identified to allow simultaneously the removal of the D2D devices interference at the level of the base station and the removal of the cellular users (CU) interference at the level of D2D devices. To optimally solve the joint channel and power allocation (PA) problem, a time-efficient solution of the PA problem in the FD context is elaborated. By means of graphical representation, the complex non-convex PA problem is efficiently solved in constant time complexity. This enables the global optimal resolution by successively solving the separate PA and channel assignment problems. The performance of the proposed strategy is compared against the classical state-of-the-art FD and HD  scenarios, where SIC is not applied between CUs and IoT devices. The results show that important gains can be achieved by applying mutual SIC NOMA in the IoT-cellular context, in either HD or FD scenarios.
	\end{abstract}
	\section{Introduction}
	\IEEEPARstart{F}{ollowing} the growth in the number of connected devices in recent years, unprecedented highs are expected for the near future \cite{cisco}. Coupled with the expected increase in data traffic and the limited available spectrum, the corresponding network densification will require novel efficient solutions to supply the ever increasing demand. Full-duplex (FD) communication combined with device-to-device (D2D) communication represent an attractive solution to leverage the challenges of future generation networks. 
	
	By enabling direct communication between nearby devices avoiding the transit through base stations (BS) or gateways \cite{d2dintro}, D2D communication leverages network resources and enables increasing the number of connected devices. In FD communication, a node can send and receive simultaneously using the same frequency resource. The achieved gain can go up to a virtual two-fold increase in spectral efficiency (SE) compared to half-duplex (HD) send-then-receive systems. In return, a self-interference (SI) is incurred due to the transmitted signal looping back into the receiver, thus limiting its appeal compared to HD to the point where this latter may even outperform FD in some cases. Nonetheless, the improvement in antenna architecture and in SI cancellation circuitry dramatically reduces the residual self-interference (RSI) \cite{SIbetter,SIbetter2,removeSI}, advocating for the use of FD for future communication standards.

	The increasing demand for higher spectral efficiency and massive connectivity for internet of things (IoT) steered research towards non-orthogonal multiple access (NOMA) techniques. The sharing of multiple devices over the same time and frequency resource enables important SE gains, lower latency communications, and increased number of connected IoT devices  \cite{iot1,JF_ICC,MJ_DAS_VTC,Zhu,ding,R11,Marie_tcom}. In power-domain NOMA, signals are differentiated in the power dimension \cite{R10}, where superposition coding of users signals is used at the transmitter, and successive interference cancellation (SIC) is performed at the receiver side. At the level of a receiver, the message with the highest power is decoded first and then subtracted from the total received signal, then the message with the second highest power is extracted and so on until the user decodes its own message. Works such \mbox{\cite{chen2019secure,cao2019secrecy,VirtualMIMO}} used NOMA and FD for cooperative relaying as well as secrecy provision, calling on to virtual MIMO as physical layer security enabler for fifth generation centric IoT applications. In this paper, we study the resource allocation problem for D2D communications systems underlaying cellular networks using the NOMA technique coupled with FD transmission scenarios. This being said, the broader problem of resource optimization in IoT could be tackled via multiple other tools such as next generation reconfigurable intelligent surfaces (RIS) \mbox{\cite{nazih}}, quantum computing inspired metaheuristics \mbox{\cite{fi3lan}}, and much more.\\
	
	\subsection{Related Works}
	Recently, attention was focused on the combination of NOMA with D2D communications in underlay mode.
	The study in \cite{accessNOMAsumthroughput} considers resource block assignment and power allocation (PA) in a downlink NOMA system with D2D. HD is used in the D2D pairs, and CUs are grouped in NOMA clusters. The influence of the HD-D2D users over the SIC decoding orders of CUs is accounted for in both the block assignment and the PA phase, because the interference they generate may change the decoding order. However, NOMA SIC is not used to decode the interfering signals of the collocated D2D pairs. The same is true for \cite{underlayD2DNOMA}, but additional power constraints are introduced on the D2D pairs to maintain the same SIC decoding orders at CUs as for the case of D2D-disabled systems. The work in \cite{d2dNomachannelAllocation} introduces the concept of D2D group, where a D2D transmitter communicates with multiple D2D receivers via NOMA. To maximize the network sum-throughput, sub-channel allocation is conducted using many-to-one matching for CU-D2D grouping and optimal PA is approximated iteratively via successive convex approximation. When limiting the number of multiplexed D2Ds to one per CU user, the work in \cite{D2DNoma} provides a joint D2D-CU grouping and PA strategy for energy efficiency maximization: the Kuhn-Munkres technique is applied successively for channel allocation, while optimal PA is obtained using the Karush-Kuhn-Tucker conditions. In all the preceding studies, NOMA is applied either between the CU users \cite{underlayD2DNOMA}, or between users of the same D2D group \cite{d2dNomachannelAllocation,D2DNoma}, but the interference cancellation of the D2D signals at the level of CU users (and inversely) is not considered. At most, attention is given towards managing the SIC decoding order at the level of the CUs in \cite{accessNOMAsumthroughput,underlayD2DNOMA}, or at the level of the D2D receivers in \cite{d2dNomachannelAllocation,D2DNoma}.

	The work in \cite{D2DhdNOMA} tackles the problem of HD-D2D throughput maximization in an uplink system where NOMA is used between D2D and CU users. If the D2D causes strong interference on the BS, its signal can be decoded then subtracted before retrieving the CU signal. However, FD-D2D is not studied and SIC occurs only at the level of the BS, i.e. not at the devices levels. Besides, the information-theoretic conditions for SIC feasibility are not considered in the study.
	In \cite{D2Dinterlay}, an efficient graph-based scheme is proposed to maximize the D2D sum-rate of an uplink system. To that end, an interlay mode is introduced to HD-D2D communication where a D2D pair can join a NOMA group to remove the interference between it and the cellular NOMA users. However, the conditions for applying SIC - and thus for determining the SIC decoding order - are only conditioned by the ascending order of channel gains between the senders and the receivers. In other words, the interfering signals that can be canceled are the ones that are attributed channel gains better than that of the useful signal, regardless of their power level at reception. This may lead to outage probabilities of one if no PA measures are taken to guarantee SIC stability as shown in \cite{ding}. The work in \cite{iot3} incorporates NOMA into D2D cellular networks to maximize system connectivity. Unlike \cite{D2Dinterlay}, the D2D NOMA-aided modes are defined according to the SIC orders at the level of the D2D and the BS. The SIC decoding orders are governed by the strong interfering signal which is bound to the channel conditions as well as the used PA. The optimal PA and mode selection are solved in the presence of decoding signal-to-interference-plus-noise ratios (SINRs) threshold constraints, then the user pairing problem is turned into a min-cost max-flow problem which is solved by the Ford–Fulkerson algorithm. However, the case of FD-D2D NOMA-aided network was not addressed neither in this study, nor in the entire literature combining NOMA and D2D. \commentaire{se souvenir de mentionner dans le manuscrit que ces etudes ne determinent pas les conditions d'application du mutual SIC \ldots ils se contentent uniquement d'ecrire les conditions de SINR sans les developper jusqu'au bout.}
	
	\subsection{Contributions }
	In this paper,  we study the combination of NOMA with HD-D2D and FD-D2D systems using mutual SIC. We introduced the concept of mutual SIC in our previous works \cite{JA_VehTech,AK_HDAS,ref:MutSIC} where we showed that the signals of two or more users multiplexed in NOMA, and powered by distributed antennas, can be decoded and removed at the level of every user in the NOMA cluster. In a D2D cellular scenario, this translates into removing the interference of the D2D devices at the level of the BS, and the interference of CU users at the level of D2D pairs. The objective of the paper is to maximize the D2D sum-throughput, through joint optimal channel and power allocation, while maintaining the quality of service (QoS) requirement for all CU users. The main contributions of the paper are summarized as follows:

	\commentaire{
		In our previous works \cite{JA_VehTech,AK_HDAS,ref:MutSIC}, we introduced the concept of mutual SIC, where we showed that the signals of two or more users multiplexed in NOMA, and powered by distributed antennas, can be decoded and removed at the level of every user in the NOMA cluster. In this paper,  we study the combination of mutual SIC NOMA with HD-D2D and FD-D2D systems, in order to remove the interference of the D2D devices at the level of the BS, and the interference of CU users at the level of D2D pairs. The objective of the paper is to maximize the D2D sum-throughput, through joint optimal channel and power allocation, while maintaining the quality of service (QoS) requirement for all CU users. The main contributions of the paper are summarized as follows: }
	\begin{itemize}
		\item We derive the power multiplexing conditions (PMC) and SIC conditions allowing for the interference cancellation between D2D and CU users. The SIC constraints are the set of conditions that make mutual SIC feasible from the information theory perspective, i.e. the conditions on achievable rates at the respective levels of the users. The PMCs are the set of conditions that make the mutual SIC technique feasible from a practical implementation perspective, i.e. they guarantee that the power level of the message to be decoded, $m_d$, at a given SIC iteration, is greater than the sum of the remaining messages, so that $m_d$ can be detected and discerned from the noise plus background interference.  This guarantees SIC stability, since every signal is ensured to be the dominant signal during its decoding \cite{Zhu}\cite{ding}.
		
		\item We show that the PMCs imply the SIC conditions for both HD and FD transmission modes, which greatly reduces the PA problem complexity for the case of FD-SIC. 
		
		\item We solve analytically the PA problem for all the transmission strategies, especially for the case of FD-SIC where an efficient procedure is provided to optimally solve the D2D rate maximization problem with constant time complexity.
		\item We show that the optimal solution of the joint PA and channel allocation problem can be achieved by successively resolving the PA problem and then the channel allocation problem. 
	\end{itemize}
	The remainder of the paper is organized as follows: section \ref{SystemModel} presents the system model and formulates the joint channel and power allocation problem, decomposing the resource allocation into separate PA and channel allocation problems. The PA problems of FD and HD without SIC (FD-NoSIC, HD-NoSIC) are solved in section \ref{sec:PAresolution}, while the PA problem with SIC is reformulated for HD and FD (HD-SIC, FD-SIC) in section \ref{sec:PASIC}. Mutual SIC PA is solved for the case of HD transmission in section \ref{HD-SIC}. In sections \ref{FDSICconditions}, \ref{FD-SIC}, and \ref{Sec:FD-SIC} the conditions of mutual SIC for FD-D2D are derived, the problem constraint reduction is performed, and then the proposed geometrical resolution is exposed, allowing for a cost-effective resolution of the FD-SIC PA problem. The channel allocation is discussed in section \ref{Munkures}.  Simulation results are presented in section \ref{numResults}, and conclusions are drawn in section \ref{sec:conclusion}.
	
	\section{System Model}\label{SystemModel}
	\begin{figure}[h]
		\centering
		\includegraphics[scale=0.45]{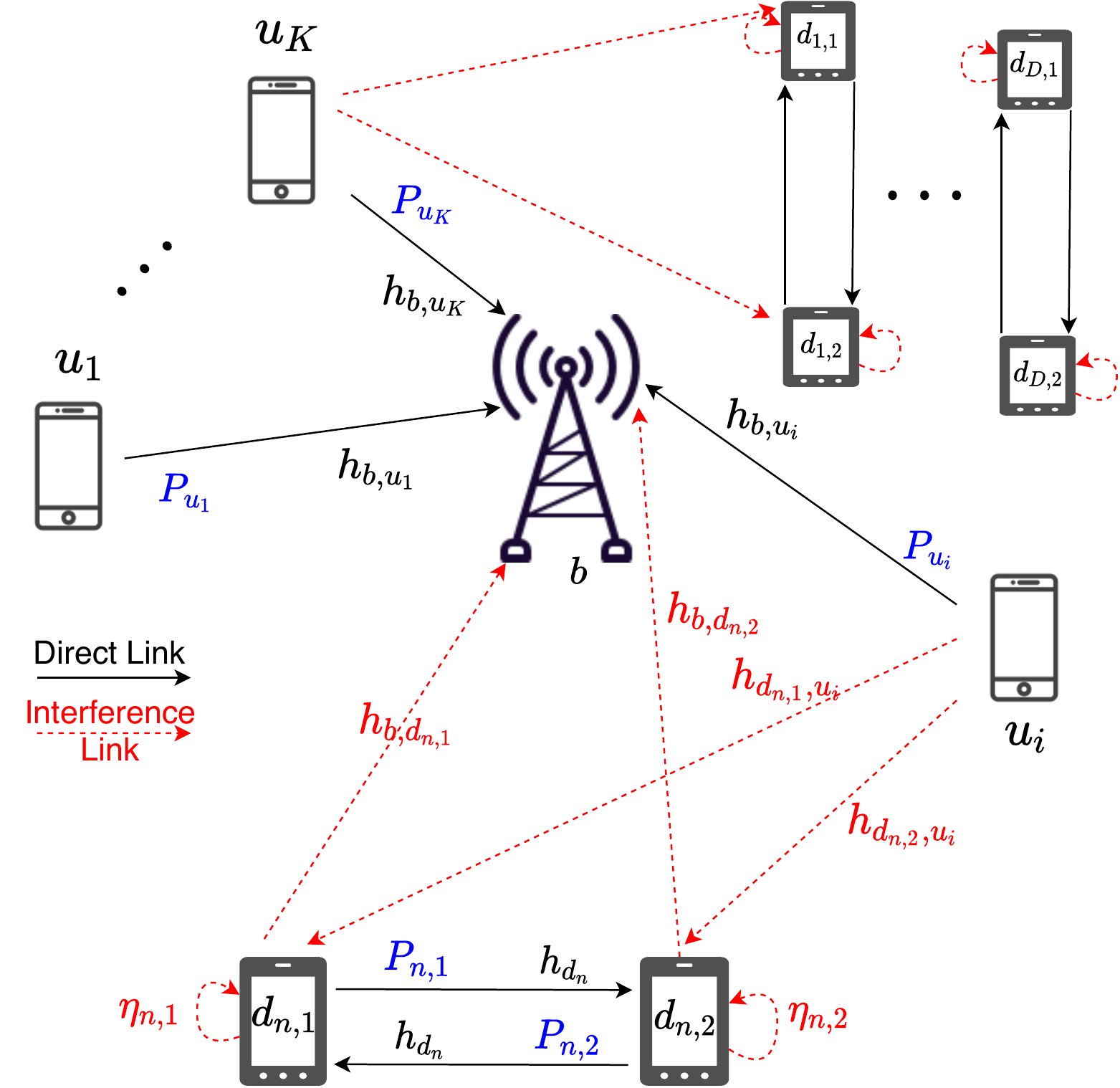}
		\caption{IoT system with $D$ D2D pairs underlaying a cellular network with $K$ CU users. \commentaire{FD inband underlay communication sharing the UL resource of a CU.} }
		\label{fig:SystemModel}
	\end{figure}

	The interference channel gains between a CU $u_i$, on the one hand, and $d_{n,1}$ and $d_{n,2}$ on the other hand, are denoted by $h_{d_{n,1},u_i}$ and $h_{d_{n,2},u_i}$ respectively. The direct link between the CU $u_i$ and the BS $b$ has a squared channel gain denoted by $h_{b,u_i}$. The message $m_{u_i}$, transmitted by $u_i$ with power $P_{u_i}$, reaches the BS with a power level $P_{u_i} h_{b,u_i}$, and causes an interference level of $P_{u_i} h_{d_{n,1},u_i}$ and $P_{u_i} h_{d_{n,2},u_i}$ at $d_{n,1}$ and $d_{n,2}$ respectively. Each device $d_{n,j}$ ($j\in \{1,2\}$) of the $n^{\text{th}}$ D2D pair can transmit a message $m_{n,j}$ of power $P_{n,j}$ to the other D2D user and suffers from both the interference of user $u_i$ and its RSI power $\eta_{n,j} P_{n,j}$, with $\eta_{n,j}$ denoting the SI cancellation capability. The D2D inter-user channel gain is denoted by $h_{d_n}$ and the interference channel gains from $d_{n,1}$ and $d_{n,2}$ to the BS are denoted by $h_{b,d_{n,1}}$ and $h_{b,d_{n,2}}$ respectively.  In this study, a frequency-non-selective channel is assumed, so that the channel gains are independent from the sub-band frequency and account only for large-scale fading including path-loss and shadowing. Table \mbox{\ref{tab:Notation_Table}} contains the main notations used in the paper.
	
	\begin{table}[]
\setlength\tabcolsep{2.4pt}
\centering
\caption{Notation Table}
\label{tab:Notation_Table}
\begin{tabular}{|l|l|l|l|}
\hline
$K$ & Total number of CUs & $d_{max}$ & \begin{tabular}[c]{@{}l@{}}Maximum distance\\  between D2Ds\end{tabular} \\ \hline
$N$ & \begin{tabular}[c]{@{}l@{}}Number of  uplink  \\ channels\end{tabular} & \begin{tabular}[c]{@{}l@{}}$R_{u,min}$,\\ $P_{u,min}$\end{tabular} & \begin{tabular}[c]{@{}l@{}}Minimum required rate \\ and power for CU $u$\end{tabular} \\ \hline
$\mathbb{C}$ & Set of CUs & $\eta$ & SI cancelation factor \\ \hline
$\mathbb{D}$ & Set of D2D pairs & $\sigma^2$ & White noise power \\ \hline
$O$ & \begin{tabular}[c]{@{}l@{}}D2D chanel allocation\\  matrix\end{tabular} & \begin{tabular}[c]{@{}l@{}}$P_{1,M}$, \\ $P_{2,M}$,\\  $P_{u,M}$\end{tabular} & \begin{tabular}[c]{@{}l@{}}Maximum transmit \\ power  of  $d_1$, $d_2$\\  and $u$ resp.\end{tabular} \\ \hline
$\mathcal{R}_{D2D}$ & \begin{tabular}[c]{@{}l@{}}Maximum achievable \\ D2D rate for every\\ D2D-CU couple\end{tabular} & \begin{tabular}[c]{@{}l@{}}$\mathcal{PL}_1, \mathcal{PL}_2$\\ $\mathcal{PL}_3, \mathcal{PL}_4$\end{tabular} & \begin{tabular}[c]{@{}l@{}}Planes associated to  \\ $PMC_1$, $PMC_2$, \\ $PMC_3,$ and $PMC_4$\\  resp.\end{tabular} \\ \hline
\end{tabular}
\end{table}

	In this work, it is assumed that, prior to resource allocation and data exchange, a D2D discovery phase \cite{DirectAndNetworkAssistedDiscovery} takes place in the system, during which the D2D devices inform the BS about their desire to initiate a D2D link, and forward to the BS their estimates of the D2D-CU links ($h_{d_{n,1},u_i},h_{d_{n,2},u_i}$), as well as the D2D links ($h_{d_n}$). Therefore, the BS is assumed to have perfect knowledge of the long-term evolution of the different channel gains, through signaling exchange between the different entities. The BS then performs resource allocation based on these estimated channel gains to optimally pair the D2Ds to CUs and to instruct D2D-CU pairs of the required transmit powers on their collocated channels, according to the selected transmission scenario.

	\subsection{Formulation of the Joint Channel and Power Allocation Problem}\label{ProbFormulation}
\begin{figure*}[h]
	    \centering
	    \includegraphics[width=\textwidth]{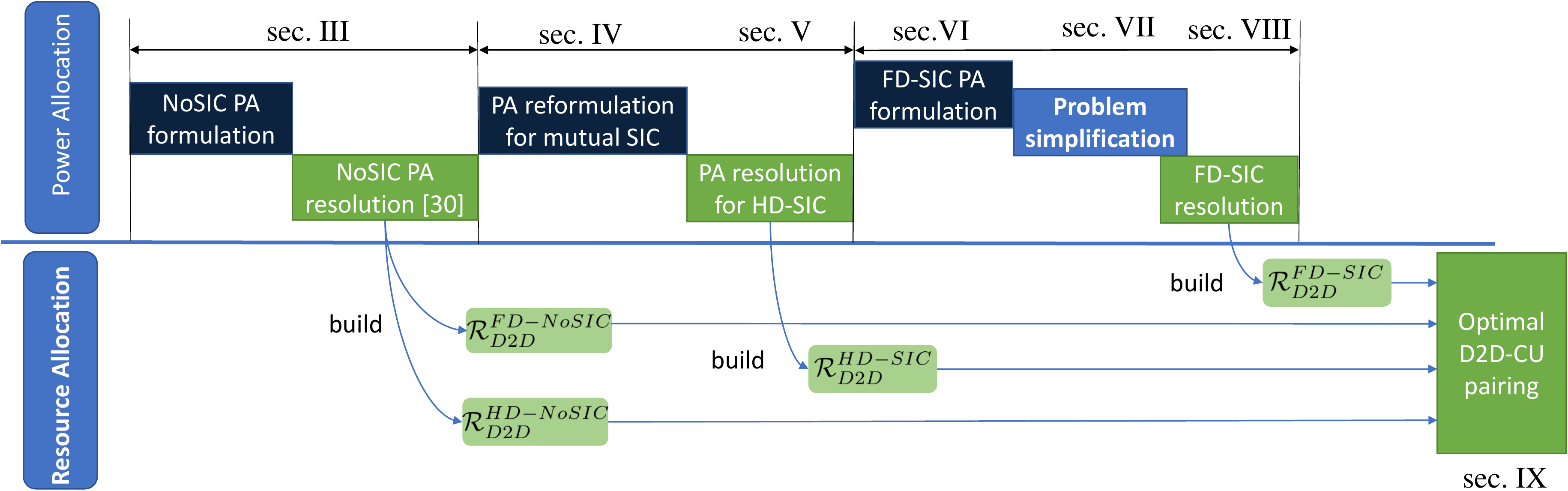}
	    \caption{Paper workflow for the resolution of the D2D channel and power allocation problem.}
	    \label{fig:workflow}
	\end{figure*}

	Let $O$ be the channel allocation matrix, with the element $o(n,i)$ at the $n^{\text{th}}$ row and $i^{\text{th}}$ column equaling one if D2D pair $n$ is collocated with CU $u_i$ and zero otherwise. Also, let $\mathcal{R}_{D2D}(n,i)$ be the maximum achievable D2D rate of the pair $n$ when collocated  with $u_i$.  Channel allocation is performed such that a D2D pair is multiplexed over a single UL channel, on the one hand, and such that a maximum of one D2D pair is multiplexed over a UL channel, on the other. The joint channel and power allocation problem for the maximization of the total D2D throughput can be cast as: 
	\begin{gather}
	\max_{\{O, P_{n,1},P_{n,2},P_{u_i}\}} \left(\sum_{i=1}^{K}\sum_{n=1}^{D} o(n,i)\times \mathcal{R}_{D2D}(n,i)\right) \nonumber
	\\
	\text{s.t. }\sum_{i=1}^{K} o(n,i) = 1,\forall n \in \{1,\ldots, D\},  
	\sum_{n=1}^{D} o(n,i) \leq 1, \forall i \in \{1,\ldots, K\} \label{MunkresConditions}
	\end{gather} 
	with $\mathcal{R}_{{D2D}}(n,i)$ the solution to:
	\begin{subequations}\label{genericProblem}
		\begin{equation}
		\mathop{\max}_{\{P_{n,1},P_{n,2},P_{u_i}\}} R_{D2D}(n,i), \tag{\ref{genericProblem}}
		\end{equation}
		\begin{multicols}{2}
			\noindent
			\begin{align}
			\text{s.t. }\quad &R_{u_i} \geq R_{u_i,min},  \label{contRu} \\
			&P_{n,1} \leq P_{n,1,M}, \label{contP1}
			\end{align}
			\noindent
			\begin{align}
			P_{u_i} &\leq P_{u_i,M}, \label{contPu}\\
			P_{n,2} &\leq P_{n,2,M},  \label{contP2}
			\end{align}
		\end{multicols}
	\end{subequations}
	\vspace{-0.45cm}
	\noindent with $R_{u_i,min}$ the minimum target rate of $u_i$, $R_{u_i}$ its achieved rate, and $R_{D2D}(n,i)$ the D2D rate, sum of achieved rates by $d_{n,1}$ ($R_{d_{n,1}}$) and $d_{n,2}$ ($R_{d_{n,2}}$). The $M$ in the subscripts of $P_{u_i,M}, P_{n,1,M},P_{n,2,M}$ refers to the maximum transmit powers of $u_i,d_{n,1}$ and $d_{n,2}$ respectively.\\
	From the structure of Problem \eqref{MunkresConditions}, and since CU users are allocated orthogonal channels, the performance of a given D2D-CU pair is independent from the network activity over the remaining channels in the system.  
	Therefore, one can optimize the throughput of all possible D2D-CU pairs, constructing a $D\times K$ table of achievable rates, and then proceed to the optimal allocation of channels to D2Ds in a second phase, thus pairing D2Ds to CUs based on their achievable rate to maximize the D2D sum-throughput in the system. The aim of the following sections is to obtain the optimal PAs of the four transmission methods FD-NoSIC, HD-NoSIC, HD-SIC and FD-SIC in order to build their corresponding tables of achievable rates $\mathcal{R}_{D2D}^{FD-NoSIC},\mathcal{R}_{D2D}^{HD-NoSIC}, \mathcal{R}_{D2D}^{HD-SIC}$, and $\mathcal{R}_{D2D}^{FD-SIC}$ respectively. Based on these tables, the optimal channel allocation is conducted in Section \ref{Munkures}. The overall procedure spanning from the PA problems formulations in section \mbox{\ref{sec:PAresolution}}, to the optimal channel assignment in section \mbox{\ref{Munkures}}, is depicted in Fig. \mbox{\ref{fig:workflow}}.

	\section{Power Allocation for No-SIC Scenarios}\label{sec:PAresolution}
	From hereinafter, since the optimal D2D rate of all $(n,i)$ couples is to be computed and because the resolution of the PAs is independent of the elected D2D-CU couple, we drop the indices relative to a specific D2D pair and CU user. Hence, user $u$ designates the CU user at hand, and $d_1$ and $d_2$ are the corresponding D2D pair. The involved channels gains are therefore denoted as $h_d$, $h_{b,d_1}$, $h_{b,d_2}$, $h_{d_1,u}$, $h_{d_2,u}$ and $h_{b,u}$, and the transmit powers of $d_1,d_2$ and $u$ are $P_1, P_2, P_u$, with their power limits $P_{1,M},P_{2,M},P_{u,M}$.
	\subsection{FD-NoSIC}\label{FD-NoSIC}
	In FD, $d_1$ and $d_2$ transmit simultaneously, thus they both suffer from RSI. Since, in this method, SIC is not attempted at the levels of $d_1,d_2$ and the BS, the SINRs at the level of the BS and the D2D users are given by: 
	\begin{gather}
	SINR_b = \frac{P_u h_{b,u}}{P_1 h_{b,d_1} + P_2h_{b,d_2} + \sigma^2}\nonumber\\
	SINR_{d_1} = \frac{P_2 h_d}{P_u h_{d_1,u} + \eta_1 P_1 + \sigma^2}, SINR_{d_2} = \frac{P_1 h_d}{P_u h_{d_2,u} + \eta_2 P_2 + \sigma^2}, \label{FD-NoSIC-SINR} 
	\end{gather}
	with $\sigma^2$ the additive Gaussian noise power. The achieved rates are expressed according to the Shannon capacity theorem \commentaire{\revised{ajouter une ref}}: 
	\begin{gather}
	R_u = B \log_2 ( 1 + SINR_b),  \label{equ:FDCuRate}\\
	R_{d_1} = B \log_2 ( 1 + SINR_{d_1}),  \quad  R_{d_2}=B\log_2 ( 1 + SINR_{d_2}), \label{equ:FDD2DRate}
	\end{gather}
	with $B$ the bandwidth of each UL channel resource. Due to the interference terms in (\ref{FD-NoSIC-SINR}), Problem \eqref{genericProblem} is non-convex. To solve it, a geometrical representation can be used, leading to the analytical global solution in \cite{Gallen}. This method is adopted in our work to derive the results of the FD-NoSIC scenario in the performance assessment section. 
	\subsection{HD-NoSIC}\label{HD-NoSIC}
	The time slot is now divided into two equal half-time slots where $d_1$ and $d_2$ alternately transmit and receive information. To maximize the total D2D rate, the optimization is conducted separately in the two half-time slots. In the first half, $d_1$ transmits information ($P_2=0$). In Problem \eqref{genericProblem}, the objective function and CU rate are now: 
	\begin{gather*}
	R_{D2D,1}= R_{d_2} = B\log_2(1 + \frac{P_1 h_d}{P_{u,1}h_{d_2,u} + \sigma^2}),\\
	R_{u,1} = B \log_2 ( 1 + \frac{P_{u,1} h_{b,u}}{P_1 h_{b,d_1}+\sigma^2}).
	\end{gather*}
	Also, Problem \eqref{genericProblem} is constrained only by \cref{contRu,contPu,contP1}.
	Note that $P_{u,1}$ is the transmit power of $u$ during the first half-time slot. $R_{D2D,1}$ is strictly increasing with $P_1$ and decreasing with $P_{u,1}$; therefore, to maximize $R_{D2D,1}= R_{d_2}$, $P_1$ should be increased and $P_{u,1}$ decreased as long as $R_{u,1}$ satisfies the minimum rate condition of the CU. Consequently, $P_1$ should be increased as much as possible and then $P_{u,1}$ is obtained as a function of $P_1$ ($P_{u,1} = f(P_1)$) by enforcing an equality between $R_{u,1}$ and $R_{u,min}$. If for $P_1=P_{1,M}$, $f(P_{1,M})\leq P_{u,M}$, the couple $(P_{1,M}, f(P_{1,M}))$ is retained as the ($P_1$, $P_{u,1}$) solution; otherwise, the couple $(f^{-1}(P_{u,M}),P_{u,M})$ delivers the best solution. The same reasoning is applied for the second half-time slot (where $P_1=0$) to maximize $R_{D2D,2} = R_{d_1}$. The total user $u$ and D2D rates are given by: 
	\begin{gather*}
	R_u = \frac{1}{2} R_{u,1} + \frac{1}{2} R_{u,2},\quad   
	R_{D2D} = \frac{1}{2} R_{d_1} + \frac{1}{2} R_{d_2}.
	\end{gather*}
	
	\section{PA Problem Modification for HD and FD with mutual SIC (HD-SIC and FD-SIC)}\label{sec:PASIC}
	Using a SIC receiver at the level of the BS and the D2D users, interfering messages can be decoded then subtracted from the received message, canceling thereby the interference in both FD and HD scenarios. Let $m_1$ and $m_2$ be the messages transmitted by the devices $d_1$ and $d_2$. In the case of FD, the BS can decode and subtract successively $m_1$ then $m_2$, or inversely, before proceeding to the decoding of $m_u$ (the message transmitted by the CU); hence, two decoding orders are possible. Users $d_1$ and $d_2$ can also remove the interference of $u$, leading to the following SINR expressions: 
	\begin{gather*}
	SINR_{d_1} = \frac{P_2 h_d}{\eta_1 P_1 + \sigma^2},
	SINR_{d_2} = \frac{P_1 h_d}{\eta_2 P_2 + \sigma^2},
	SINR_b = \frac{P_u h_{b,u}}{\sigma^2}.
	\end{gather*}
	The SINRs are replaced in \cref{equ:FDCuRate,equ:FDD2DRate} to  obtain $R_u$ and $R_{D2D}= R_{d_1} + R_{d_2}$ that will be used in Problem \eqref{genericProblem}. For the case of HD, the SINRs in the first half-time slot are: 
	\begin{gather*}
	SINR_{d_2} = {P_1 h_{d}}/{\sigma^2}, \qquad \qquad  SINR_{b} = {P_u h_{b,u}}/{\sigma^2}.
	\end{gather*}
	In the second half-time slot, $SINR_{b}$ is the same and $SINR_{d_1} = P_2 h_{d} / \sigma^2$. Problem \eqref{genericProblem} is now reformulated in each time slot by expressing the rates using the present SINRs. However, additional constraints relative to the SIC feasibility must be added to the problem. They are derived in the next sections for HD-SIC, then for FD-SIC. Afterwards, Problem \eqref{genericProblem} is solved for the two SIC scenarios.
	
	\section{PA for HD-SIC scenario}\label{HD-SIC}
	Consider the first half-time slot, where $u$ and $d_1$ are transmitting and $b$ and $d_2$ are receiving. Hereafter, we develop the mutual SIC constraints between $b$ and $d_2$ (as a receiver). Let $SINR_i^{m_j}$ be the SINR of the message $m_j$ at the level of user $i$ ($i$ is either $d_1$, $d_2$ or $b$, and $j$ is either $1$, $2$ or $u$). For $b$ to successfully decode the message $m_1$ transmitted by $d_1$ to $d_2$, the received rate of $m_1$ at the level of $b$ must be greater than the rate of $m_1$ at the level of $d_2$. Thus, we must have: $SINR_{b}^{m_1} > SINR_{d_2}^{m_1}$. Similarly, the rate condition for the decoding of $m_u$ at the level of $d_2$ is derived from the condition $SINR_{d_2}^{m_u} > SINR_{b}^{m_u}$. This situation is equivalent to the case of two different radio resource heads (RRHs) transmitting both messages to two separate receivers, which was studied in \cite{JA_VehTech}. It was shown that the SINR conditions lead to: \begin{equation}
	h_{b,d_1}h_{d_2,u} > h_{d}h_{b,u}. \label{necessary1}
	\end{equation}
	In addition to condition \eqref{necessary1}, the PMCs must be verified, in order to ensure that the message to be decoded first at the level of a receiver has a higher power level than that of the remaining message \cite{ref:MutSIC}. The PMCs for the decoding of $m_u$ and $m_1$ at the level of $d_2$ and $b$ are given by: 
	\begin{gather}
	\begin{rcases}
	P_{u,1} h_{d_2,u} > P_1 h_{d}\\
	P_1 h_{b,d_1} > P_{u,1} h_{b,u}
	\end{rcases}\Rightarrow A = \frac{h_d}{h_{d_2,u}} < \frac{P_{u,1}}{P_1} < \frac{h_{b,d_1}}{h_{b,u}} = B. \label{hdpmc}
	\end{gather}
	Note that \eqref{necessary1} is satisfied if \eqref{hdpmc} is satisfied, since \eqref{necessary1} is equivalent to $A<B$. Therefore, the PMCs encompass the rate conditions while being more restrictive. Problem \eqref{genericProblem} now only includes the additional constraint \eqref{hdpmc} for the first time slot. The HD-SIC rate expressions are as follows:	
	\begin{equation*}
	R_{D2D,1} = R_{d_2} = B\log_2 (1 + \frac{P_1 h_{d}}{\sigma^2}),
	R_{u,1} = B\log_2(1+\frac{P_{u,1} h_{b,u}}{\sigma^2}).    
	\end{equation*}	
	
	Maximizing $R_{d_2}$ lies in the increase of $P_1$. Also, guaranteeing the CU rate $R_{u,min}$ can be achieved by setting $P_{u,1}$ to $P_{u,m}=(2^{\frac{R_{u,min}}{B}}-1)\sigma^2/h_{b,u}$. However, due to the PMCs, the increase in $P_1$ is very likely to increase $P_{u,1}$ according to the range of allowed values in \eqref{hdpmc}, leading to an excess of CU rate. Since maximization of network throughput (i.e. sum of D2D and CU rates) is not the objective of this study, we select from the range of admissible $R_{u,1}$ values, the one closest to $R_{u,min}$. With that criterion in mind, the power allocation problem for D2D rate maximization is solved by increasing $P_1$ as much as possible (possibly until $P_{1,M}$) and adjusting $P_{u,1}$ accordingly. 
	\begin{figure}[h]
		\includegraphics[scale=0.75]{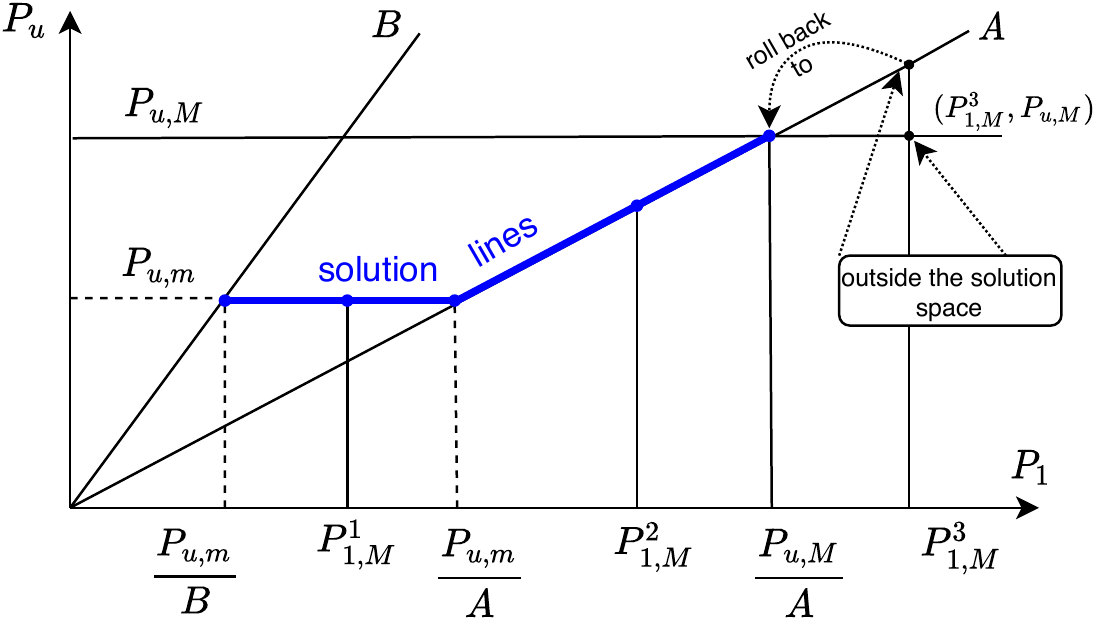}
		\caption{Schematic of the solution space to the HD-SIC PA problem, for different $P_{1,M}$ values.}
		\label{fig:optimalPA}
	\end{figure}
	The proposed PA procedure, illustrated in Fig. \ref{fig:optimalPA}, operates as follows: if $P_{1,M} < P_{u,m}/A$, keep the couple $(P_1 = P_{1,M}, P_{u,1} = P_{u,m})$. This case is represented by the example $P^1_{1,M}$ on the horizontal blue line in Fig. \ref{fig:optimalPA}. If this is not the case, check if $A P_{1,M} > P_{u,M}$. If yes  (cf. example $P^3_{1,M}$ in Fig. \ref{fig:optimalPA}), the solution is $({P_{u,M}}/{A},P_{u,M})$; if not (cf. example $P^2_{1,M}$), the solution is $(P_{1,M},AP_{1,M})$. Restricting the solution space to the blue lines in Fig. \ref{fig:optimalPA} guarantees that the CU always transmits at the minimum necessary power that respects the problem constraints.
	Note that if $P_{1,M}$ is too low ($<P_{u,m}/B$), the problem is not feasible even when \eqref{necessary1} is verified. \\
	For the second time slot, the same methodology is followed, where the PMCs and the new necessary and sufficient channel conditions are given by: 
	\begin{gather}
	h_{d_1,u}h_{b,d_2} > h_{b,u}h_d, \label{necessary2}\\
	A^{'}=\frac{h_d}{h_{d_1,u}}< \frac{P_{u,2}}{P_2}< \frac{h_{b,d_2}}{h_{b,u}} = B^{'}.\label{hdpmc2}
	\end{gather}
	As a conclusion, in the HD-SIC scenario, the system checks for the validity of the channel condition corresponding to the half-time slot before going through the procedure described above. If the channel condition is not favorable or if no solution exists (i.e. $P_{u,m} > P_{u,M}$ or $P_{1,M}< P_{u,m}/B$ for the first half, and $P_{2,M}< P_{u,m}/B^{'}$ for the second half), the system reverts to the HD-NoSIC solution of \cref{HD-NoSIC}. This leads to four combinations of SIC/NoSIC procedures, two for every half-time slot, and they are all included in the HD-SIC algorithm.
	
	\section{Derivation of the SIC conditions for FD mutual SIC}\label{FDSICconditions}
	
	In this scenario, we are looking for the conditions that allow $d_1$ to decode $m_u$, $d_2$ to decode $m_u$, and $b$ to decode $m_1$ and $m_2$. As already mentioned, two decoding orders are possible at the level of $b$. 
	\subsection{First decoding order: $b$ decodes $m_2$ then $m_1$}\label{sec:FirstDecodingOrder}
	We first start by studying the mutual SIC constraints between $b$ and $d_1$ (as a receiver). For $b$ to successfully decode the message $m_2$ transmitted by $d_2$ to $d_1$, we must have:
	\begin{gather*}
	SINR_b^{m_2} > SINR_{d_1}^{m_2},\\
	\frac{P_2 h_{b,d_2}}{\sigma^2 + P_1h_{b,d_1} + P_uh_{b,u}} > 
	\frac{P_2 h_{d}}{\sigma^2 + P_1\eta_1 + P_u h_{d_1,u}}.
	\end{gather*}
	Since practical systems are interference-limited \cite{InterferenceLimit1,InterferenceLimit2}, the noise power is negligible compared to the interfering terms which yields the SIC condition:
	\begin{equation}
	P_1(h_{b,d_2}\eta_1 - h_{d}h_{b,d_1}) + P_u(h_{d_1,u}h_{b,d_2} - h_{d}h_{b,u})>0.  \label{sicS2atB}
	\end{equation}
	In addition to condition \eqref{sicS2atB}, the PMCs must be verified. Since $b$ decodes $m_2$ first, then we have the following PMC for the decoding of $m_2$:
	\begin{equation}
	P_2 h_{b,d_2} > P_{1}h_{b,d_1} + P_{u}h_{b,u}. \label{pmcS2atB}
	\end{equation}
	For $d_1$ to be able to remove the interference of $m_u$ prior to retrieving $m_2$, we must have $SINR_{d_1}^{m_u} > SINR_{b}^{m_u} $, which leads to:
	\begin{gather}
	P_1(h_{d_1,u}h_{b,d_1} \!-\! h_{b,u}\eta_1)\! + \!P_2(h_{d_1,u}h_{b,d_2} \!-\! h_{b,u}h_{d})>0, \label{sicSuatD1}
	\end{gather}
	and the corresponding PMC is: 
	\begin{equation}
	P_{u}h_{d_1,u} > P_{2}h_{d} + P_{1}\eta_1. \label{pmcSuatD1}
	\end{equation}
	Regarding the mutual SIC between the receivers $b$ and $d_2$, the decoding of $m_1$ at the level of $b$ requires $	SINR_{b}^{m_1}$ to be greater than $SINR_{d_2}^{m_1}$:
	\begin{gather}
	\frac{P_{1}h_{b,d_1}}{\sigma^2 + P_uh_{b,u}}
	>
	\frac{P_1h_{d}}{\sigma^2 + P_2\eta_2 + P_u h_{d_2,u}}, \nonumber\\
	P_2 h_{b,d_1}\eta_2 > P_u (h_{b,u}h_{d} - h_{d_2,u}h_{b,d_1}).  \label{sicS1atB}
	\end{gather}
	Note that $SINR_{b}^{m_1}$ does not include $P_2$ since $m_2$ is decoded and canceled prior to $m_1$. The corresponding PMC is given by:
	\begin{equation}
	P_1 h_{b,d_1} > P_u h_{b,u}. \label{pmcS1atB}
	\end{equation}
	At the level of $d_2$, $SINR_{d_2}^{m_u}$ must be greater than $SINR_{b}^{m_u}$ to decode and subtract $m_u$ before retrieving $m_1$. This yields the following condition:
	\begin{equation}
	P_1 (h_{b,d_1}h_{d_2,u} - h_{d}h_{b,u}) > P_2 \eta_2 h_{b,u} \label{sicSuatD2}
	\end{equation}
	Finally, the PMC at the level of $d_2$ is given by:
	\begin{gather}
	P_u h_{d_2,u} > P_{1}h_{d} + P_2 \eta_2 \label{pmcSuatD2}
	\end{gather}
	\subsection{Second decoding order: $b$ decodes $m_1$ then $m_2$}
	Following the same reasoning as in section \ref{sec:FirstDecodingOrder}, for the case where $m_1$ is decoded before $m_2$ at the level of $b$, the PMC and rate constraints for a full SIC between $d_1$ and $b$, and $d_2$ and $b$, are obtained and listed below: 
	\begin{gather}
	P_1\eta_1 h_{b,d_2} > P_u (h_{b,u}h_{d} - h_{b,d_2}h_{b,d_1})  \label{FirstConditionDO2} \\
	P_2 (h_{d_1,u}h_{b,d_2} - h_{u,b}h_{d})> h_{u,b}\eta_1 P_1\\
	P_2 (h_{b,d_1}\eta_2 \!- \!h_{b,d_2}h_{d}) \!+\! P_u (h_{b,d_2} h_{b,d_1} \!-\! h_{b,u}h_{d})\!>\!0\\
	P_1 (h_{d_2,u}h_{b,d_1} \!-\! h_{d}h_{u,b}) \!+\! P_2  (h_{d_2b}h_{d_2,u} \!-\! h_{u,b}\eta_2)\!>\!0\\
	P_{2}h_{b,d_2} > P_{u}h_{b,u} \label{debut}\\
	P_uh_{d_1,u} > P_{2}h_{d} + P_{1}\eta_1 \label{2}\\
	P_1h_{b,d_1} > P_{u}h_{b,u} + P_{2}h_{b,d_2}  \label{3}\\
	P_uh_{d_2,u} > P_{1}h_{d} + P_{2}\eta_2 \label{LastConditionDO2}
	\end{gather}
	
	In addition to constraints \cref{contP1,contP2,contPu,contRu}, Problem \eqref{genericProblem} now includes eight new constraints that express the full SIC feasibility (either equations \eqref{sicS2atB} to \eqref{pmcSuatD2} or \eqref{FirstConditionDO2} to \eqref{LastConditionDO2}, depending on the decoding order). Solving this optimization problem with inequality constraints by means of the standard Karush–Kuhn–Tucker conditions implies exploring all the possible combinations of active/inactive constraints (an inequality constraint is active if it is verified with equality). This results in a total of $2^{12} -1$ combinations to be considered. To reduce this exorbitant complexity, the interplay between SIC rate conditions and PMCs is analyzed in the next section, targeting the removal of redundant constraints. 
	
	\section{PA Problem Simplification of FD-SIC by Constraint Reduction}\label{FD-SIC}
	Consider the first decoding order at the level of $b$ where $m_2$ is decoded before $m_1$. The PMCs for the decoding of $m_1$ at the level of $b$ and of $m_u$ at the level of $d_2$ are given by \eqref{pmcS1atB} and \eqref{pmcSuatD2}. By multiplying $\eqref{pmcS1atB}$ by $h_{d_2,u}$ and adding it to $\eqref{pmcSuatD2}$ multiplied by $h_{b,u}$, one can eliminate $P_u$ to obtain:
	
	\begin{equation*}
	P_1 (h_{b,d_1}h_{d_2,u} - h_{d}h_{b,u}) > P_2 \eta_2 h_{b,u},
	\end{equation*}
	which is the SIC condition $\eqref{sicSuatD2}$ introduced to remove $m_u$ at the level of $d_2$. Also, eliminating $P_1$ from the two PMCs by means of adding $\eqref{pmcS1atB}$ multiplied by $h_{d}$ to $\eqref{pmcSuatD2}$ multiplied by $h_{b,d_1}$ yields \eqref{sicS1atB}. Consequently, the PMCs for the decoding of $m_1$ at the level of $b$, and $m_u$ at the level of $d_2$ imply their counterpart rate conditions. Moreover, it is noted from \eqref{sicSuatD2} that the same necessary condition \eqref{necessary1} that is found in HD-SIC between $b$ and $d_2$ as receivers, is obtained for the application of FD-SIC between $d_2$ and $b$:
	\begin{equation}
	h_{b,d_1}h_{d_2,u} > h_{d}h_{b,u}. \tag{\ref{necessary1}}
	\end{equation}
	Note that if \eqref{necessary1} is not true, \eqref{sicSuatD2} becomes impossible to satisfy no matter $P_1$ and $P_2$; however, when \eqref{necessary1} is true, \eqref{sicSuatD2} can be satisfied under an adequate power play between $P_1$ and $P_2$.
	
	We now move to the PMC and SIC conditions for the decoding of $m_2$ and $m_u$ at the level of $b$ and $d_1$ respectively, i.e. \eqref{pmcS2atB}, \eqref{pmcSuatD1}, \eqref{sicS2atB} and \eqref{sicSuatD1}.  
	By adding \eqref{pmcS2atB} multiplied by $h_{d}$ to \eqref{pmcSuatD1} multiplied by $h_{b,d_2}$, $P_2$ is eliminated to yield:
	\begin{equation}
	P_u (h_{d_1,u}h_{b,d_2} - h_{b,u} h_{d}) > P_1  (h_{b,d_1} h_{d} + \eta_1 h_{b,d_2}), \label{hint}
	\end{equation} which can be further transformed into:
	\begin{gather*}
	P_1(\eta_1 h_{b,d_2}- h_{b,d_1} h_{d}) + P_u (h_{d_1,u}h_{b,d_2} - h_{b,u} h_{d}) > 2 P_1  \eta_1 h_{b,d_2}  \nonumber \\
	\Rightarrow P_1(\eta_1 h_{b,d_2}- h_{b,d_1} h_{d}) + P_u (h_{d_1,u}h_{b,d_2} - h_{b,u} h_{d}) > 0. 
	\end{gather*}
	Thus, the PMCs \eqref{pmcS2atB} and \eqref{pmcSuatD1} imply \eqref{sicS2atB}. In fact, not only do they imply the rate condition, but it is clear that the PMCs represent more restrictive constraints than rate conditions. Finally, eliminating $P_u$ from the PMCs through the combination of \eqref{pmcS2atB} multiplied by $h_{d_1,u}$ with \eqref{pmcSuatD1} multiplied by $h_{b,u}$ yields:
	\begin{gather}
	P_2 (h_{b,d_2} h_{d_1,u} - h_d h_{b,u}) > P_1 ( h_{b,d_1}h_{d_1,u} + \eta_1 h_{b,u}), 
	\label{hint2}
	\end{gather}
	which can be rearranged into: 
	\begin{gather*}
	P_2(h_{d_1,u}h_{b,d_2} \!-\! h_{b,u}h_{d})\! +\! P_1(h_{d_1,u}h_{b,d_1}\! -\! h_{b,u}\eta_1)\!\! >\!\! 2 P_1 h_{b,d_1}h_{d_1,u} \nonumber\\
	\Rightarrow \eqref{sicSuatD1}.
	\end{gather*}
	Once again, the PMCs for the decoding of $m_2$ and $m_u$ at $b$ and $d_1$ imply their rate condition counterparts. Note that the necessary channel condition that appears from \cref{hint,hint2} is the same as in the case of HD-SIC in the second half-time slot:
	\begin{equation}
	h_{d_1,u}h_{b,d_2} > h_{b,u}h_d.  \tag{\ref{necessary2}}
	\end{equation}
	Also, the combinations of \eqref{pmcS1atB} with \eqref{pmcSuatD1}, and \eqref{pmcSuatD2} with \eqref{pmcS2atB}, while eliminating $P_u$, give the following conditions: 
	\begin{gather}
	P_1 (h_{b,d_1} h_{d_1,u} - \eta_1h_{b,u}) > P_2 h_{d}h_{b,u},  \nonumber \\
	P_2 (h_{b,d_2} h_{d_2,u} - \eta_2 h_{b,u}) > P_1 ( h_{b,d_1}h_{d_2,u} + h_dh_{b,u}). \nonumber
	\end{gather}
	These inequalities yield two other necessary, but not sufficient, channel conditions for the application of full SIC to the system: 
	\begin{gather}
	h_{b,d_1} h_{d_1,u} > \eta_1h_{b,u}, \label{necessary3}\\
	h_{b,d_2} h_{d_2,u} > \eta_2 h_{b,u}. \label{necessary4}
	\end{gather}
	
	Repeating the same procedure for the second decoding order delivers the same results: 1) the PMCs encompass the rate conditions, 2) the same necessary four channel conditions \eqref{necessary1}, \eqref{necessary2}, \eqref{necessary3}, and \eqref{necessary4} are obtained. 
	
	Therefore, in the FD-SIC scenario, the system checks the validity of \cref{necessary1,necessary2,necessary3,necessary4} prior to solving the PA problem for each decoding order. If the channel conditions are not valid or no solution is obtained for \eqref{genericProblem}, the FD-SIC algorithm reverts to the FD-NoSIC procedure described in section \ref{FD-NoSIC}. As a conclusion for this section, Problem \eqref{genericProblem} is now only equipped with the PMC set corresponding to the decoding order (i.e. \cref{pmcS1atB,pmcS2atB,pmcSuatD1,pmcSuatD2}, or \cref{debut,2,3,LastConditionDO2}), in addition to constraints \cref{contP1,contP2,contPu,contRu}. This reduces the number of combinations of active/inactive constraints from $2^{12}-1$ to $2^{8}-1$ which is still considerable. The aim of the next section is to workaround the need of a full search over the corresponding $255$ cases for determining the optimal PA. This is done by efficiently determining the meaningful constraint combinations, based on the geometrical interpretation of the FD-SIC PA problem. Considerable complexity reductions arise from this approach as shown next.

	\section{Solution for FD-SIC Optimal PA}
	\label{Sec:FD-SIC}
	The proposed geometrical resolution of the FD-SIC D2D rate maximization PA problem is presented in detail for the first decoding order. First, the geometrical representation of the solution space satisfying the PMC and power limit constraints is provided. Then, a procedure is elaborated leading to the reduction of the search space to the minimum required. Afterwards, the optimization is conducted on the resulting reduced search space. At last, a quick summary of the optimal PA procedure is presented including the required changes to obtain the optimal PA for the second decoding order.
	
	\subsection{3D Solution Space Representation}\label{sec:SolutionRepresentation}
	The four PMCs that must be satisfied in eq. \eqref{pmcS2atB}, \eqref{pmcSuatD1}, \eqref{pmcS1atB} and \eqref{pmcSuatD2} are written in the following form:
	\begin{align*}
	&P_{u}h_{b,u} < P_2 h_{b,d_2} - P_{1}h_{b,d_1}   & (PMC_1)\\
	&P_{u}h_{d_1,u} > P_{2}h_{d} + P_{1}\eta_1 & (PMC_2)\\
	&P_u h_{b,u} < P_1 h_{b,d_1}  & (PMC_3)\\
	&P_u h_{d_2,u} > P_{1}h_{d} + P_2 \eta_2 & (PMC_4)
	\end{align*}
	In the 3D space of axes $x,y,z$ representing variables $P_1, P_2$ and $P_u$ respectively, we introduce the planes $\mathcal{PL}_1, \mathcal{PL}_2, \mathcal{PL}_3$ and $\mathcal{PL}_4$ whose equation is given by the PMCs 1, 2, 3 and 4 when the conditions are met with equality. \textcolor{black}{In the following, we refer to $\mathcal{PL}_i$ as the plane \textit{derived from}, or equivalently, \textit{corresponding to}, or simply, as \textit{the plane of} $PMC_i$}. Each PMC restricts the search space either to the half space below its corresponding plane like for $PMC_1$ and $PMC_3$, or to the half space above its corresponding plane as for $PMC_2$ and $PMC_4$. On the other hand, the transmit power limits restrict the search space to the region within the parallelepiped defined by the sides $x=P_{1,M}$, $x=0$, $y=P_{2,M}$, $y=0$, $z=P_{u,M}$, $z=P_{u,m}$.
	\begin{figure}[h]
		\centering
		\includegraphics[width=0.85\linewidth]{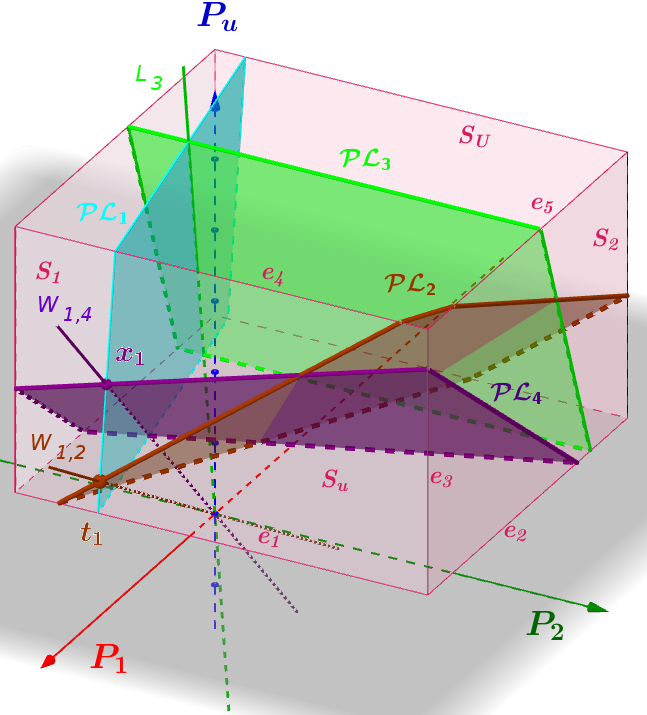}
		\caption{Schematic of the search space formed inside the intersection of the PMC planes with the parallelepiped of power limits. }\label{fig:SolutionSpace}
	\end{figure}
	To have a non-empty search space (i.e. FD-SIC is feasible), the pentahedron defined by the space region above $\mathcal{PL}_2$ and $\mathcal{PL}_4$ and below $\mathcal{PL}_1$ and $\mathcal{PL}_3$ must be non-empty, and it must have a common region with the parallelepiped.
	\begin{itemize}
		\item \underline{Non-empty pentahedron:} The pentahedron is non-empty if the intersection lines of $\mathcal{PL}_1$ with $\mathcal{PL}_2$ ($\triangleq W_{1,2}$) and $\mathcal{PL}_4$ ($\triangleq W_{1,4}$) are below the intersection line of $\mathcal{PL}_1$ with $\mathcal{PL}_3$ ($\triangleq L_3$), as shown in Fig. \ref{fig:SolutionSpace}.
		Let $\vec{u}$ be the direction vector of $W_{1,2}$; we must have $\dfrac{z(\vec{u})}{x(\vec{u})}< h_{b,d_1}/h_{b,u}$, which yields: 
		\begin{gather}
		\vec{u}=\begin{pmatrix}
		h_{d_1,u}h_{b,d_2} - h_{bu}h_{d}\\
		h_{bu}\eta_1 + h_{d_1,u} h_{b,d_1} \\
		\eta_1h_{b,d_2} + h_{b,d_1}h_{d}
		\end{pmatrix}
		\Rightarrow \frac{\eta_1h_{b,d_2} + h_{b,d_1}h_{d}}{h_{d_1,u}h_{b,d_2} - h_{bu}h_{d}}< \frac{h_{b,d_1}}{h_{b,u}}, \nonumber
		\end{gather}
		leading to the following channel condition:
		\begin{gather}
		h_{b,d_1} h_{d_1,u} - \eta_1h_{b,u} > 2h_{b,u}h_{d}\frac{h_{b,d_1}}{h_{b,d_2}}.  \label{equ:sufficientPMC2}
		\end{gather}
		Doing the same for $W_{1,4}$ with respect to $L_3$, we get the following channel condition:
		\begin{equation}
		h_{b,d_1} h_{d_2,u} - h_{b,u} h_d >  2h_{b,u} \eta_2 \frac{  h_{b,d_1}}{h_{b,d_2}}.\label{equ:sufficientPMC4}
		\end{equation}
		Note that \eqref{equ:sufficientPMC2} and \eqref{equ:sufficientPMC4} are more restrictive than the necessary conditions of eq. \eqref{necessary3} and  \eqref{necessary4}, which is normal since they turn them into sufficient channel conditions. 
		\item \underline{Pentahedron $\cap$ parallelepiped:} For the pentahedron to have a non-empty intersection with the parallelepiped, it is sufficient to make sure that the intersection line of $\mathcal{PL}_1$ with $\mathcal{PL}_3$ ($L_3$) intersects the plane $z = P_{u,m}$ within the $P_{1,M}$ and $P_{2,M}$ limits. These conditions on the $x,y$ coordinates of $\mathcal{PL}_3 \cap \mathcal{PL}_1 \cap P_{u,m}$ yield the constraints:
		\begin{equation}
		P_{u,m}\frac{h_{b,u}}{ h_{b,d_1}} < P_{1,M} \quad \&\& \quad  
		2P_{u,m}\frac{h_{b,u}}{h_{b,d_2}} < P_{2,M} \label{equ:Pmax}.
		\end{equation}
	\end{itemize}
	Conditions \eqref{equ:sufficientPMC2}, \eqref{equ:sufficientPMC4} and \eqref{equ:Pmax} form the necessary and sufficient constraints for the existence of a solution to the FD-SIC PA problem according to the first decoding order.
	\subsection{Search Space Reduction}\label{sec:SpaceReduction}
	We prove in this section that the optimal solution lies on 
	the intersection line of $\mathcal{PL}_2,\mathcal{PL}_4$ or the lower side of the parallelepiped $S_{u}$, with one of the outer sides of the parallelepiped $S_1, S_2, S_{U}$ (cf. Fig. \ref{fig:SolutionSpace}), respectively defined by:\\ $x=P_{1,M} \text{ for } (y,z)\in [0,P_{2,M}]\times [P_{u,m},P_{u,M}], y=P_{2,M} \text{ for }(x,z)\in [0,P_{1,M}]\times [P_{u,m},P_{u,M}]$, and $ z=P_{u,M} \text{ for }(x,y)\in[0,P_{1,M}]\times[0,P_{2,M}]$.
	\begin{proposition}
		The optimal solution lies on one of the outer sides of the parallelepiped.
	\end{proposition}
	\begin{proof}
		The D2D rate is given by: 
		\begin{gather*}
		R_{D2D}(P_1,P_2) = B\log_2 ( 1 + \frac{P_1 h_d}{P_2\eta_2 + \sigma^2}) + B\log_2 ( 1 + \frac{P_2h_d}{P_1\eta_1 + \sigma^2})
		\end{gather*}
		For any couple $(P_1,P_2)$, and $\forall \beta > 1$, the throughput of $(\beta P_1,\beta P_2)$ is greater than $R_{D2D}(P_1,P_2)$ since: 
		\begin{align*}
		R_{D2D}&(\beta P_1, \beta P_2) =\\
		& B\log_2 ( 1 + \frac{P_1 h_d}{P_2\eta_2 + \sigma^2/\beta}) + B\log_2 ( 1 + \frac{P_2h_d}{P_1\eta_1 + \sigma^2/\beta})
		\\
		&> B\log_2 ( 1 + \frac{P_1 h_d}{P_2\eta_2 + \sigma^2}) + B\log_2 ( 1 + \frac{P_2h_d}{P_1\eta_1 + \sigma^2})
		\\& = R_{D2D}(P_1,P_2)
		\end{align*}
		Therefore, given an initial triplet $(P_1,P_2,P_u)$, a higher throughput-achieving triplet can be obtained by simply multiplying the components by a factor larger than 1. The higher the $\beta$, the higher the throughput, meaning that $\beta$ should be increased until reaching the boundaries of the region, which can be either $P_{1,M}, P_{2,M}$ or $P_{u,M}$.
	\end{proof}	 
	Moreover, it is clear that the D2D rate is independent of $P_u$. This means that when moving on a vertical line in the solution space, $R_{D2D}$ is constant and $P_u$ only affects the CU rate. To keep the CU rate as close as possible to $R_{u,min}$, we select the smallest $P_u$ value from the range of admissible values for a given $(P_1,P_2)$ couple. Since every point in the solution space must be on top of $\mathcal{PL}_2$ and $\mathcal{PL}_4$, the minimum allowed value of $P_u$ is given by forcing the equality either on $PMC_2$ or on $PMC_4$, according to the one that delivers the higher minimum value of $P_u$  for the considered $(P_1,P_2)$ couple.
	
	As a conclusion, the optimal solution lies on the intersection segment of one of the outer sides of the parallelepiped $S_1$, $S_2$, or $S_{U}$, with one of the planes $\mathcal{PL}_2$, $\mathcal{PL}_4$, or $S_{u}$, resulting in a total of eight possible combinations (eight segments). Given the shape of the solution space, some of the combinations are mutually exclusive. The aim of the next section is to determine which subset of segments should be accounted for in the power optimization process, depending on the channel conditions of the D2D-CU couple. \commentaire{Depending on the shape of the solution space, some of the combinations are mutually exclusive. The aim of the next section is to determine which combinations (i.e. which segments) should be accounted for to find the optimal solution.}

	\subsection{Selection of the Useful Intersections}\label{sec:UsefulIntersections}
\begin{figure}[h]
		\centering
		\includegraphics[width=0.95\linewidth]{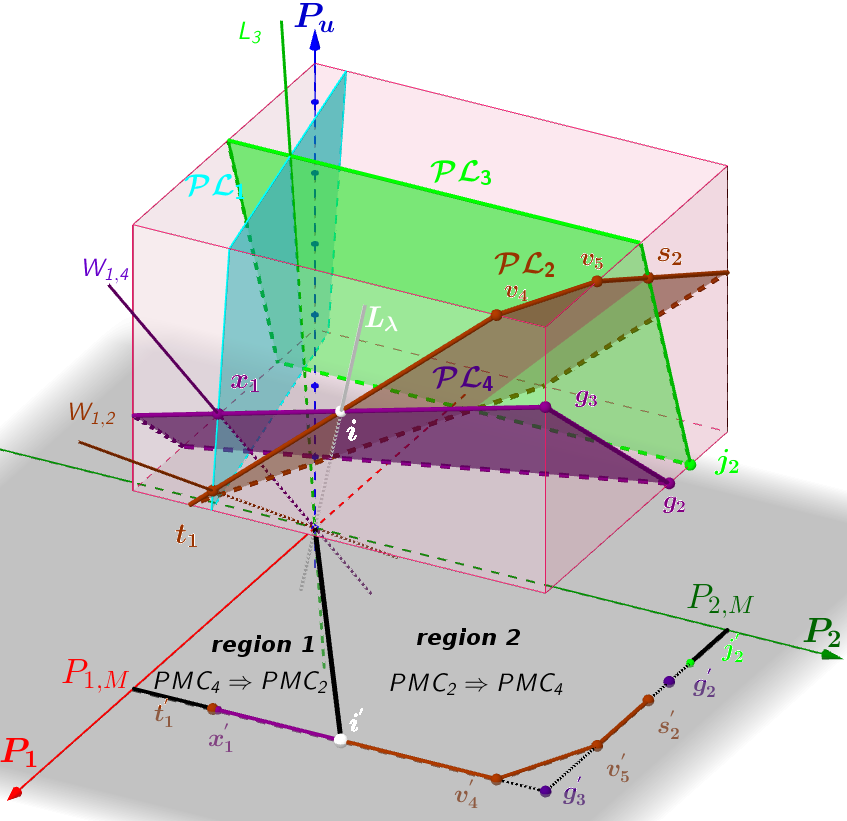}
		\caption{Schematic of the solution space showing the regions of dominance of $PMC_4$ over $PMC_2$ and vice-versa. }\label{fig:PMCregions}
	\end{figure}

\begin{figure}[h]
		\centering
		\begin{subfigure}{0.98\linewidth/2}
			\includegraphics[width=\textwidth]{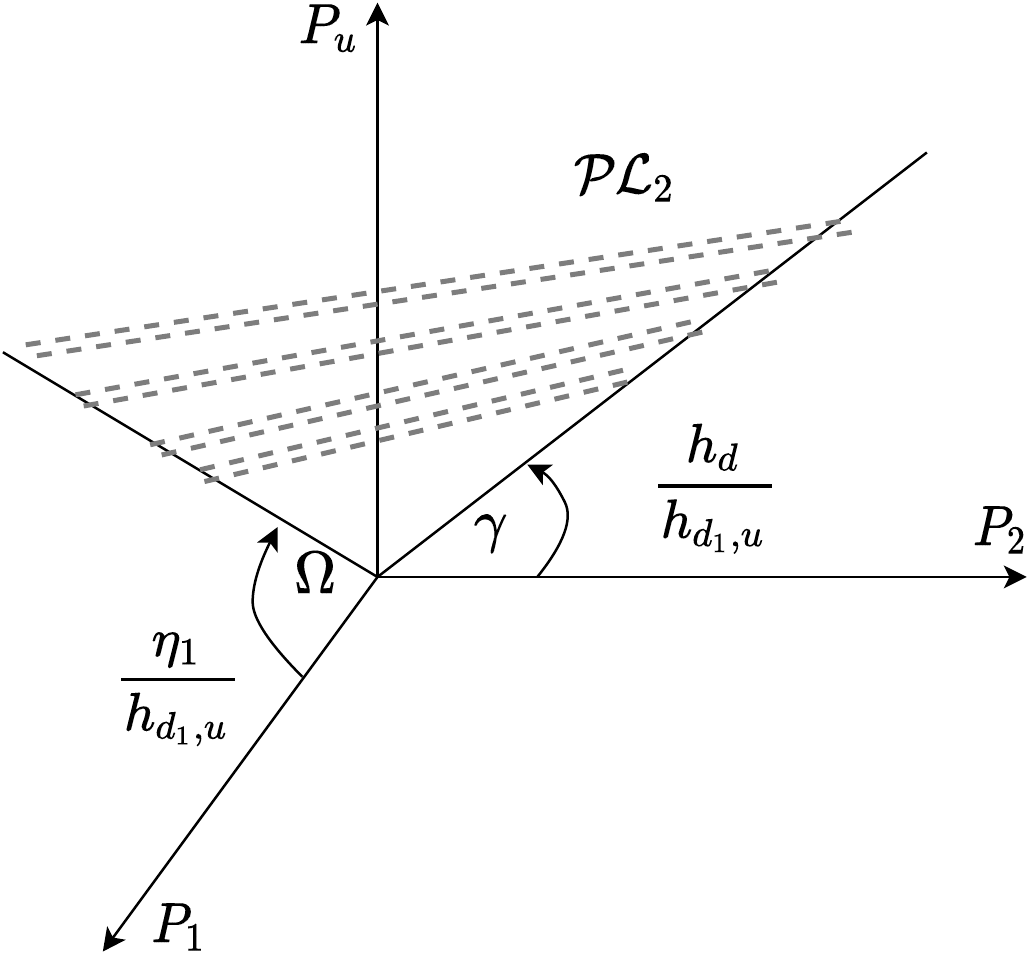}
			\caption{$ PMC_2$}\label{fig:PMC2}
		\end{subfigure}
		\begin{subfigure}{0.99\linewidth/2}
			\includegraphics[width=\textwidth]{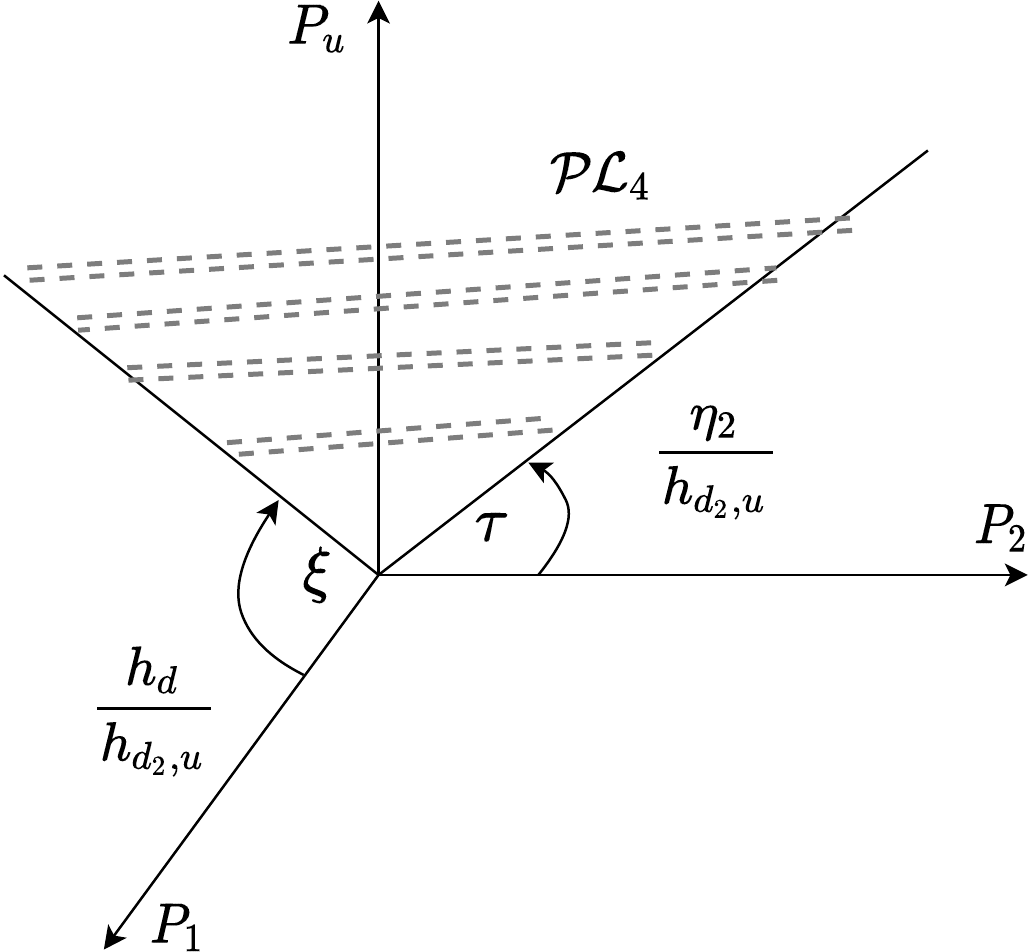}
			\caption{$PMC_4$}
			\label{fig:PMC4}
		\end{subfigure}
		\caption{Isolated schematics of $\mathcal{PL}_2$ and $\mathcal{PL}_4$ in the 3D space.}\label{fig:PMC24cases}
	\end{figure}

	As can be seen from Fig. \ref{fig:SolutionSpace}, some of the eight intersections can be discarded. For example, the intersection of $S_{u}$ with $S_1$ and $S_2$ is not relevant, since the value of $P_u$ is decided by $PMC_2$ and $PMC_4$, whose planes are on top of $S_{u}$ near the sides $S_1$ and $S_2$. Fig. \ref{fig:PMCregions} shows the projection on the $(P_1,P_2)$ plane of the partition of the space into two vertical regions where $PMC_4$ encompasses $PMC_2$ for region 1, and $PMC_2$ encompasses $PMC_4$ for region 2. The plane separating the two regions is the vertical plane passing through the straight line $L_{\lambda} \triangleq \mathcal{PL}_4 \cap \mathcal{PL}_2$.
	Therefore, for the case of Figs. \ref{fig:SolutionSpace} and \ref{fig:PMCregions}, the D2D rate optimization is to be conducted over the segment $\overline{x_1 i} \cup \overline{i v_4}$ which is included in $S_1$, over the segment $\overline{v_4 v_5}$ included in $S_{U}$, and on the segment $\overline{v_5 s_2}$ over $S_{2}$.  By doing so, the optimization over the segments $\overline{t_1 i}$, $\overline{i g_3}$, $\overline{g_3 g_2}$ and $\overline{g_2 j_2}$ is avoided.
	\\
	Therefore, the first step in reducing the number of intersections to be considered lies in determining which of $PMC_4$ and $PMC_2$ encompasses the other, and for which region of the space. To that end, a schematic of $\mathcal{PL}_2$ and $\mathcal{PL}_4$ is presented in Figs. \ref{fig:PMC2} and \ref{fig:PMC4}, showing their intersection with the planes defined by $P_1 = 0$, and $P_2 = 0$. The angles of these intersection lines and their slopes are shown in Fig. \ref{fig:PMC24cases}.
	\subsubsection{\textbf{Interplay between $\boldsymbol{PMC_2}$ and $\boldsymbol{PMC_4}$}}\label{sec:PMC24}
\begin{figure*}[t]
		\centering
		\begin{subfigure}{0.99\linewidth/2}
			\includegraphics[width=\textwidth]{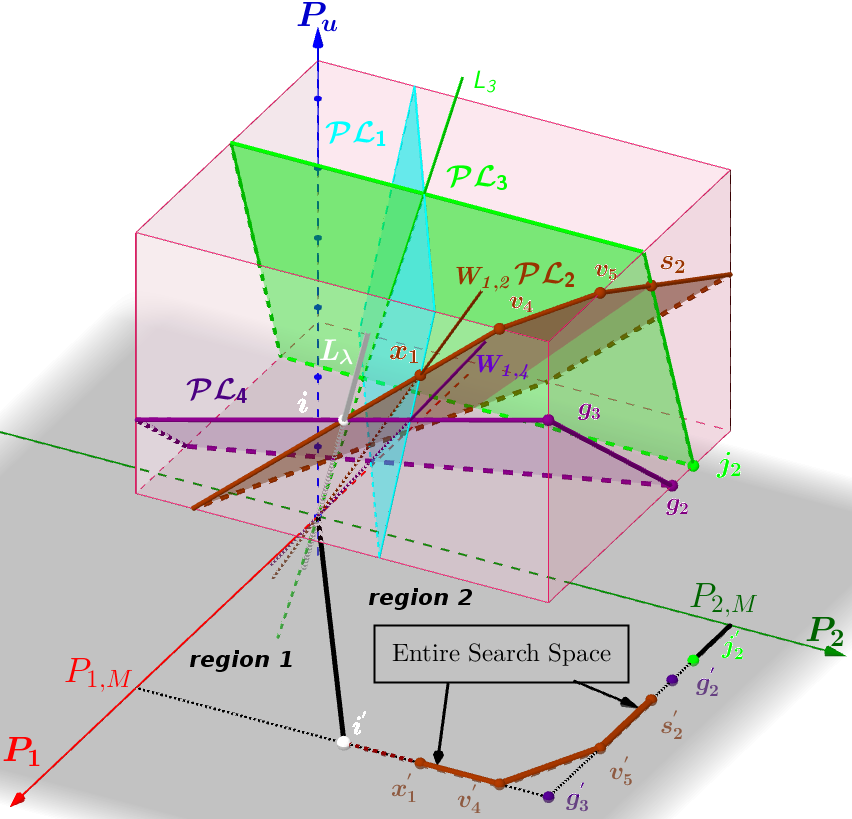}
			\caption{Search space included in region 2}\label{fig:PMC2toPMC4}
		\end{subfigure}
		\begin{subfigure}{0.98\linewidth/2}
			\includegraphics[width=\textwidth]{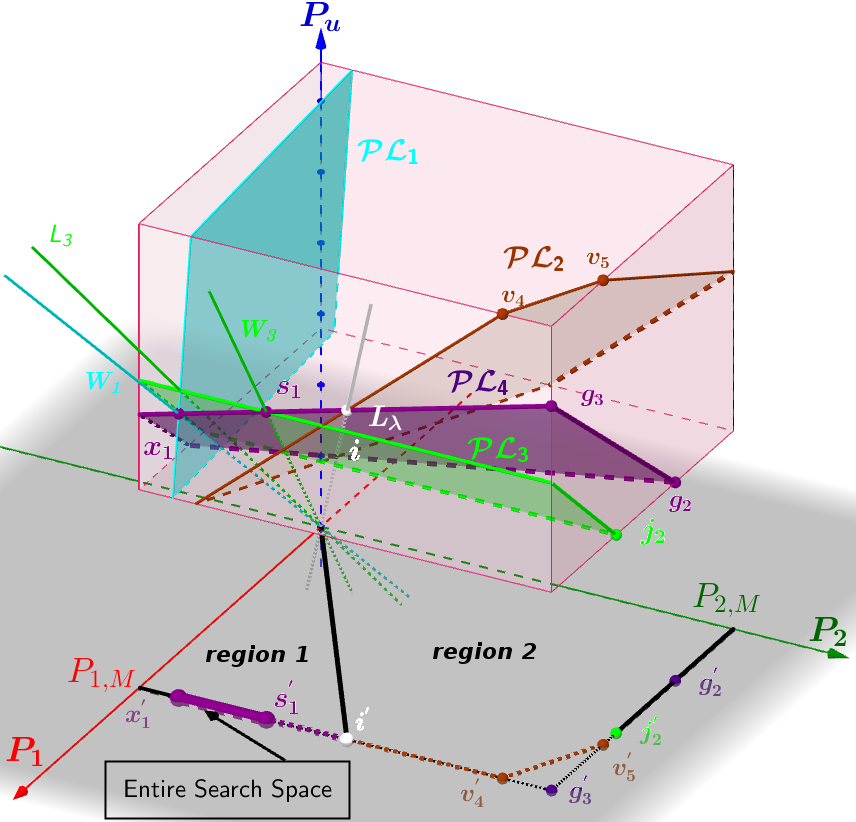}
			\caption{Search space included in region 1}\label{fig:PMC4toPMC2}
		\end{subfigure}
		\caption{Figures representing solution search spaces included in regions 1 or 2.}\label{fig:SearchSpaceIncluded}
	\end{figure*}

	Depending on the angles $\Omega, \gamma, \xi$ and $\tau$, four cases are identified to determine the interplay between $PMC_2$ and $PMC_4$:
	\begin{enumerate}
		\item $\Omega > \xi$, $\gamma > \tau$: $PMC_2$ encompasses $PMC_4$ ($PMC_2 \Rightarrow PMC_4$) over all the positive $(P_1,P_2)$ plane.
		\item $\Omega < \xi$, $\gamma < \tau$: $PMC_4$ encompasses $PMC_2$ ($PMC_4 \Rightarrow PMC_2$) over all the positive $(P_1,P_2)$ plane.
		\item $\Omega < \xi$, $\gamma > \tau$: $PMC_4$ encompasses $PMC_2$ in region 1 and $PMC_2$ encompasses $PMC_4$ in region 2, (cf. Fig. \ref{fig:PMCregions}). 
		\item $\Omega > \xi$,  $\gamma < \tau$: $PMC_4$ encompasses $PMC_2$ in region 2 and $PMC_2$ encompasses $PMC_4$ in region 1.
	\end{enumerate}
	Before proceeding, it must be noted that even for cases 3) and 4), a PMC may encompass the other on the entire search space if the actual search space is included either in region 1 or 2. This is depicted in the examples of Fig. \ref{fig:SearchSpaceIncluded} which take back the conditions of Fig. \ref{fig:PMCregions} with some modifications. In Fig. \ref{fig:PMC2toPMC4}, $\mathcal{PL}_1$ is such that $W_{1,4}$ is at the right side of $L_{\lambda}$ ($W_{1,4}$ is in region 2), then the search space is included in region 2 and only $PMC_2$ needs to be accounted for. The other scenario is represented in Fig. \ref{fig:PMC4toPMC2} where $\mathcal{PL}_3 \cap \mathcal{PL}_2$ is at the left side of $L_{\lambda}$ (in region 1), hence $PMC_4$ encompasses $PMC_2$ over the entirety of the search space. The first scenario occurs when $L_{\lambda}$ is on top of $\mathcal{PL}_1$, and the second one occurs when $L_{\lambda}$ is on top of $\mathcal{PL}_3$. The explicit channel conditions enabling each scenario are derived in detail in Appendix \ref{app1}, where it is shown that two simple tests are required to determine if the search space is included in any region. 
	
	As a conclusion, by comparing $\Omega$ to $\xi$ and $\gamma$ to $\tau$, and following the discussion in Appendix \ref{app1}, the number of intersections to be considered is reduced by selecting the appropriate PMC between $PMC_2$ and $PMC_4$ in the corresponding space region. For the sake of clarity, we introduce $PMC_{2,4}$ as the efficient combination of $PMC_2$ and $PMC_4$, \commentaire{i.e. $PMC_{2,4} = PMC_2 \forall (P_1,P_2)/ \mathcal{PL}_2$ is on top of $\mathcal{PL}_4$ $PMC_{2,4}$ is}  given by: 
	\begin{gather*}
	P_{u} \geq
	\begin{cases}
	\dfrac{P_2 h_d + P_1 \eta_1}{h_{d_1,u} }, &\mbox{if } P_{2}(\frac{h_d}{h_{d_1,u}} -\frac{\eta_2}{h_{d_2,u}}) > P_1 (\frac{h_d}{h_{d_2,u}}  - \frac{\eta_1}{h_{d_1,u}} )\\
	\dfrac{P_1 h_d + P_2 \eta_2}{h_{d_2,u}}, &\mbox{elsewhere}.
	\end{cases}
	\end{gather*}
	\subsubsection{\textbf{Selection of the useful parallelepiped sides}}\label{sec:SideSelection}
	With $PMC_{2,4}$ at hand, the next step is to reduce the unnecessary sides of the parallelepiped, which do not intersect with $\mathcal{PL}_{2,4}$, or that do intersect with $\mathcal{PL}_{2,4}$ but not inside the range allowed between $\mathcal{PL}_1$ and $\mathcal{PL}_3$. To that end, we study $PMC_1$ and $PMC_3$ which do not affect the intersection segments (of $\mathcal{PL}_{2,4}$ with the parallelepiped sides) as such, but rather the end points of these intersection segments. A typical example is given in Fig. \ref{fig:PMC2toPMC4} where $PMC_1$ sets the end point $x_1$ from the side $S_1$, and $PMC_3$ sets the end point $s_2$ from the side $S_2$. \\
	
\begin{figure*}[h]
		\centering
		\begin{subfigure}{0.99\linewidth/3}
			\includegraphics[width=\textwidth]{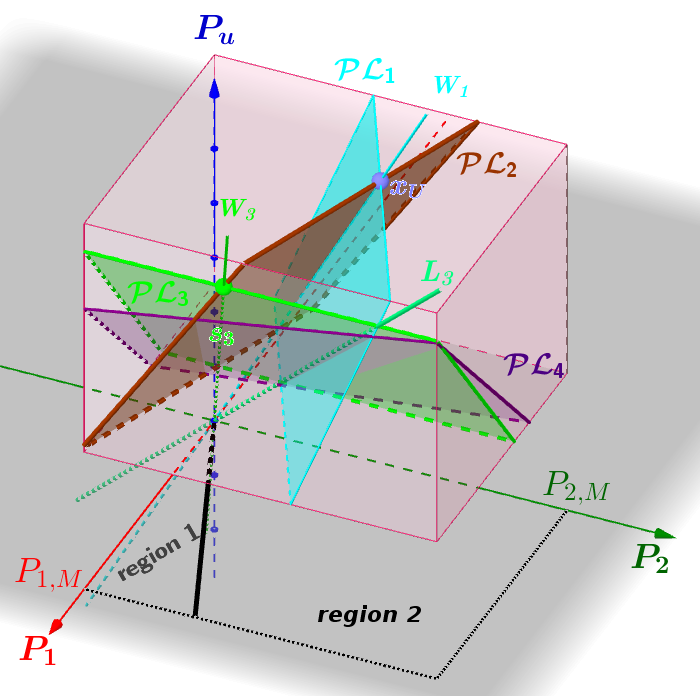}
			\caption{$ W_1 $ intercepts $S_U$ and $W_3$ intercepts $S_1$}
		\end{subfigure}
		\begin{subfigure}{0.98\linewidth/3}
			\includegraphics[width=\textwidth]{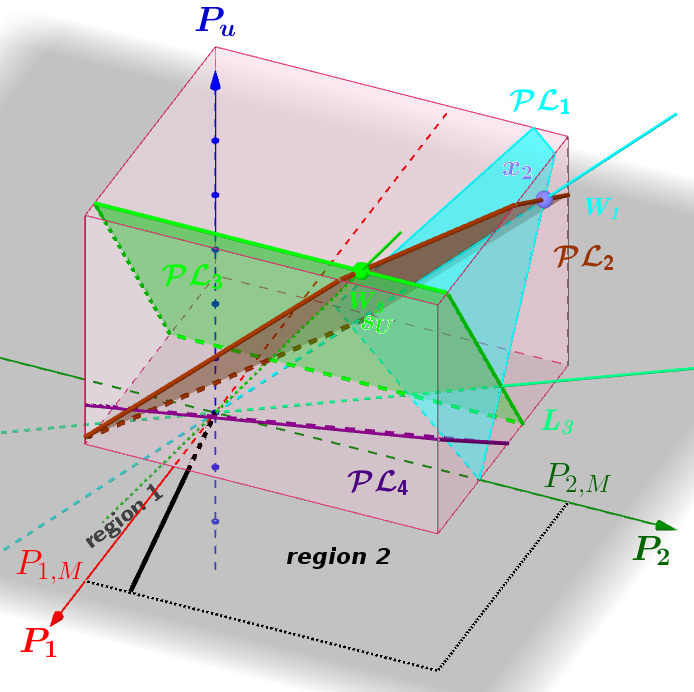}
			\caption{$ W_1 $ intercepts $S_2$ and $W_3$ intercepts $S_U$}
		\end{subfigure}
		\begin{subfigure}{0.99\linewidth/3}
			\includegraphics[width=\textwidth]{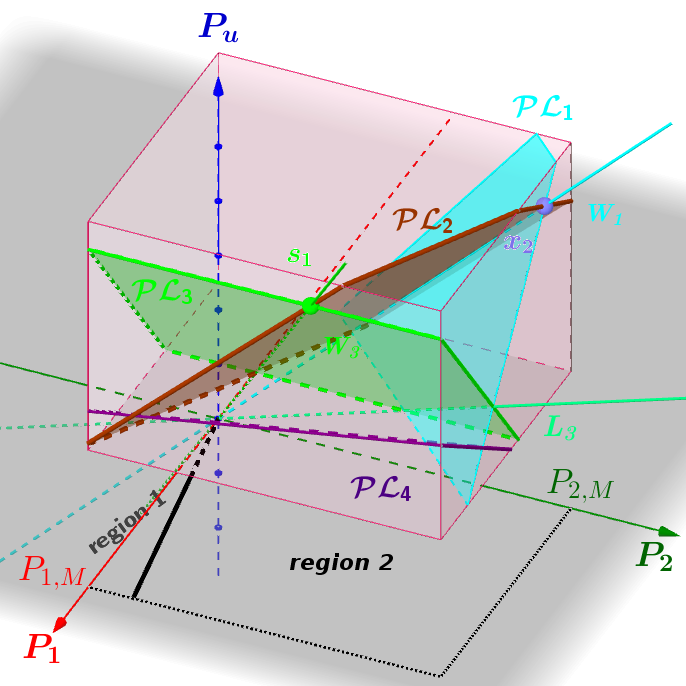}
			\caption{$ W_1 $ intercepts $S_2$ and $W_3$ intercepts $S_1$}
		\end{subfigure}
		\caption{The three non-feasible combinations between $x_i$ and $s_i$.}\label{fig:InfeasibleCases}
	\end{figure*}

	Let $W_1$ regroup the intersection lines $W_{1,2}$ and $W_{1,4}$ such that $W_1 = \mathcal{PL}_{2,4} \cap \mathcal{PL}_1$, and let $W_3$ be the intersection line of $\mathcal{PL}_3$ with $\mathcal{PL}_{2,4}$ (cf. Fig. \ref{fig:InfeasibleCases}). Each of $W_1$ and $W_3$ may intercept the sides $S_1$ or $S_2$ or $S_U$, yielding a total of nine potential combinations\commentaire{of intersection point pairs of $W_1$ with $S_1,S_2,S_U$, and $W_2$ with $S_1,S_2,S_U$}. Since each side $S_i$ is a rectangular surface within the infinite plane $\mathcal{S}_i$ of equation $P_{i} = P_{i,M}$, then $W_1$ and $W_3$ can intercept only one side of the parallelepiped ($S_{u}$ aside) for a given channel configuration. Let $x_i$ and $s_i$ be the intersection points of $W_1$ and $W_3$ with $S_i$, we have: 
	$$x_i = W_1 \cap S_i, \, s_i = W_3 \cap S_i, \forall i\in \{1,2,U\}.$$ 
	To determine which sides are hit by $W_1$ (resp. $W_3$), i.e. to determine if we have $x_1, x_2$ or $x_U$ (resp. $s_1, s_2$ or $s_U$), we consider the points $xl_i$ (resp. $sl_i$), intersections of $W_1$ (resp. $W_3$) with the planes $\mathcal{S}_i, \forall i\in \{1,2,U\}$. The coordinates of $xl_i$ are given by: 
	\begin{gather*}
	xl_1  =  \begin{pmatrix}
	P_{1,M}\\
	y(W_1 \cap \mathcal{S}_1)\\
	z(W_1 \cap \mathcal{S}_1)
	\end{pmatrix},
	xl_2 = \begin{pmatrix}
	x(W_1 \cap \mathcal{S}_2)\\
	P_{2,M}\\
	z(W_1 \cap \mathcal{S}_2)
	\end{pmatrix}, 	
	xl_U = \begin{pmatrix}
	y(W_1 \cap \mathcal{S}_U)\\
	y(W_1 \cap \mathcal{S}_U)\\
	z = P_{u,M}
	\end{pmatrix}
	\end{gather*}
	Then, two tests are needed to determine which of $x_1, x_2$ or $x_U$ occurs for the given channel states (cf. Algorithm \ref{alg:SideSelection}). 
	\begin{algorithm}
		\SetAlgoLined
		\SetKwInOut{Input}{input}\SetKwInOut{Output}{output}
		\Input{$P_{1,M}$, $P_{2,M}$, $P_{u,M}$, $P_{u,m}$, $h_{b,u}$, $h_d$,  $\eta_1$, $\eta_2$, $h_{d1,u}$, $h_{d_2,u}$, $h_{d_1,b}$, $h_{d_2,b}$}
		\KwResult{Returns $i/ W_i \cap S_i = x_i = xl_i \neq \emptyset$.}
		\caption{$W_1$ intersection with the parallalepiped}
		\eIf 
		{$y(xl_1) < P_{2,M}$}
		{
			\eIf{$z(xl_1) \leq P_{u,M}$}{$i=1$, keep $x_1$}
			{$i=U$, keep $x_U$}
		}
		{
			\eIf{$z(xl_2) < P_{u,M}$}{$i=2$, keep $x_2$}{$i=U$, keep $x_U$}
		}
		\label{alg:SideSelection}
	\end{algorithm}
	Note that $xl_i$ and $sl_i$ have positive coordinates as shown in Appendix \ref{app:EndPoints}.
	\commentaire{\textcolor{blue}{
			Note that if $y(xl_1) < P_{2,M}$ while  $z(xl_1) < P_{u,m}$, even though $xl_1\notin S_1$, we still say that $xl_i$ \textit{is on the side} $S_1$ (or on the side of $P_{1}$) and this case is associated to that of $x_i = x_1$. In this situation the face $S_1$ still hosts an optimization segment, however the endpoint previously given by  $x_1 = \mathcal{PL}_1 \cap \mathcal{PL}_{2,4} \cap S_1$ is now given by $k_1 = \mathcal{PL}_1 \cap S_u \cap S_1$. \\
			Similarly, on \textit{the side of} $P_2$, if $y(xl_1)>P_{2,M}$ while $z(xl_1) < P_{u,m}$, then the case is associated to that of $x_i=x_2$, but the point $k_2 = \mathcal{PL}_1 \cap S_u \cap S_2$ sets the segment endpoint instead of $x_2(=S_2\cap W_1=\emptyset)$}
	}
	The same tests are replicated for $s_i$. From the nine possibilities, only six combinations are actually viable because the pairs $(x_U,s_1)$, $(x_2,s_U)$ and $(x_2,s_1)$ cannot be achieved without violating \eqref{equ:sufficientPMC2} or \eqref{equ:sufficientPMC4} as can be seen in Fig. \ref{fig:InfeasibleCases}. Indeed, the three cases shown in Fig. \ref{fig:InfeasibleCases} lead to empty search spaces. The six viable pairs are given in Table \ref{tabu:segments} with the correspondence between the pairs and the parallelepiped sides hosting the useful intersection segments. 
	\begin{table}[h]
		\centering
		\begin{tabular}{c|c|c|c|}
			\cline{2-4}
			\multicolumn{1}{l|}{}                 & $S_1$        & $S_2$        & $S_U$        \\ \hline
			\multicolumn{1}{|c|}{$x_U$ and $s_U$} &              &              & $\checkmark$ \\ \hline
			\multicolumn{1}{|c|}{$x_1$ and $s_U$} & $\checkmark$ &              & $\checkmark$ \\ \hline
			\multicolumn{1}{|c|}{$x_U$ and $s_2$} &              & $\checkmark$ & $\checkmark$ \\ \hline
			\multicolumn{1}{|c|}{$x_1$ and $s_1$} & $\checkmark$ &              &              \\ \hline
			\multicolumn{1}{|c|}{$x_2$ and $s_2$} &              & $\checkmark$ &              \\ \hline
			\multicolumn{1}{|c|}{$x_1$ and $s_2$} & $\checkmark$ & $\checkmark$ & Depends		   \\ \hline
		\end{tabular}
		\caption{Table showing the sides involved in the D2D rate optimization for each of the six ($x_i$, $s_j$) viable pairs due to $PMC_1$ and $PMC_3$.} \label{tabu:segments}
	\end{table}
	Note that if $\mathcal{PL}_1$ and $\mathcal{PL}_3$ intercept $\mathcal{PL}_{2,4}$ at the same side, then the search space can be reduced to a single segment as it is the case for the first, the fourth and the fifth rows in Table \ref{tabu:segments}. For the second and third rows, two segments are involved in the D2D rate optimization. Finally, 
	in the case where $W_1$ intercepts $S_1$ and $W_3$ intercepts $S_2$ (as in Fig. \ref{fig:PMCregions}), the segment $\overline{v_4 v_5}$ belonging to $S_U$ is to be included in the D2D optimization process - in addition to the segments in $S_1$ and $S_2$ - if and only if the value of $P_u$ obtained from $\mathcal{PL}_{2,4}$ at $P_1 = P_{1,M}$ and $P_{2} = P_{2,M}$ is greater than $P_{u,M}$.

	\subsubsection{\textbf{Segments endpoints}} \label{sec:SegmentEndpoints}  Having determined the relevant intersection segments (a maximum of three segments) for the D2D rate optimization using PMCs 1 and 3, we detail hereafter how the endpoints of every segment are determined for each side of the parallelepiped. For the sake of clarity, let $e_1,e_2,e_3,e_4,e_5$ be the edges of the parallelepiped (cf. Fig. \ref{fig:SolutionSpace}) given by:
	\begin{gather*}
	e_1 = S_u \cap S_1,  e_2 = S_u \cap S_2,
	e_4 = S_U \cap S_1, e_5 = S_U \cap S_2, 
	e_3 = S_2\cap S_1
	\end{gather*} 
	Also, let the three families of points $v_i, g_i$, and $w_i$ be the intersections of $\mathcal{PL}_2, \mathcal{PL}_4$ and $\mathcal{PL}_{2,4}$ with $e_i$:
	\begin{gather*}
	v_i = \mathcal{PL}_2 \cap e_i, \quad
	g_i = \mathcal{PL}_4 \cap e_i, \quad
	w_i = \mathcal{PL}_{2,4} \cap e_i 
	\end{gather*}
	Examples of such points can be seen in Fig. \ref{fig:PMC4toPMC2} for $v_4, v_5, g_3$ and $g_2$. Note that points $w_i$ are only used to designate the points $v_i$ or $g_i$ depending on whether we are in region 1 or 2. We can now efficiently designate the segment endpoints on each side. 
	\paragraph{Side $S_2$} The optimization over $S_2$ translates into an optimization over $P_1$, since $P_2$ is equal to $P_{2,M}$. It is clear that the minimal value of $P_1$ is bound to $PMC_3$. In Fig. \ref{fig:PMCregions} for example, the minimal value of $P_1$ is obtained for the point $s_2$, intersection of $\mathcal{PL}_3$ with $\mathcal{PL}_{2,4}$. However, if $PMC_4$ were the one encompassing $PMC_2$ in region 2, the considered segment over $S_2$ would have been $\overline{g_3 g_2} \cup \overline{g_2 j_2}$; then, the minimum $P_1$ value would have been given by the point $j_2 = \mathcal{PL}_3 \cap S_2 \cap S_{u}$. Thus, we can generalize by stating that the minimal value of $P_1$ in the segment over $S_2$ is given by: 
	\begin{gather*}
	\min P_1 = \max [ x(\mathcal{PL}_3 \cap \mathcal{PL}_{2,4} \cap S_2) , x(\mathcal{PL}_3 \cap S_{u} \cap S_2) ] \nonumber \label{eq:minP1},\\
	\min P_1 = \max [ x(s_2) , x(j_2) ].
	\end{gather*}
	Regarding the maximum value of $P_1$, it can be due to the intersection of $S_2 \cap \mathcal{PL}_{2,4}$ with either $S_U$ (like for $v_5$ in Fig. \ref{fig:PMCregions}), $S_1$ (like for $g_3$), or with $\mathcal{PL}_1$ (for the case of $x_2$ and $s_2$, in the fifth row of table \ref{tabu:segments}). Also, the maximum $P_1$ value may be simply set by $k_2$, the intersection of $S_2 \cap \mathcal{PL}_1$ with $S_u$.  The maximum value of $P_1$ is given by: 
	\begin{align*}
	\max P_1 = \min [ & x(\mathcal{PL}_{2,4} \cap e_2), x(\mathcal{PL}_{2,4} \cap e_3),\\ & x(\mathcal{PL}_{2,4} \cap \mathcal{PL}_1 \cap S_2), x( \mathcal{PL}_1 \cap e_2)],
	\\
	\max P_1 = \min [  &x(w_5), x(w_3), x(x_2), x(k_2)  ]. 
	\end{align*}
	Note that for the side $S_2$, $PMC_3$ is involved in the minimum $P_1$ value, and $PMC_1$ in the maximum value.\\
	\paragraph{Side $S_1$} Regarding the side $S_1$,  $PMC_3$  is now involved in the maximum $P_2$ value, while $PMC_1$ settles the minimum $P_2$ value; its expression is given by:
	\begin{gather}
	\min P_2 = \max [ y(\mathcal{PL}_1 \cap \mathcal{PL}_{2,4} \cap S_1) , y(\mathcal{PL}_1 \cap S_{u} \cap S_1) ], \nonumber
	\\
	\min P_2 = \max [ y(x_1) , y(k_1) ]. \nonumber
	\end{gather}
	The maximum value of $P_2$ depends on which plane intercepts first $\mathcal{PL}_{2,4}$ among the three candidates: $\mathcal{PL}_3$, $S_U$ (cf. Fig. \ref{fig:PMCregions}), or $S_2$. We have: 
	\begin{gather*}
	\max P_2 = \min [  y(\mathcal{PL}_{2,4} \cap \mathcal{PL}_3 \cap S_1), 
	y(\mathcal{PL}_{2,4} \cap S_2 \cap S_1),\\ 
	y(\mathcal{PL}_{2,4} \cap S_U \cap S_1)],\\
	\max P_2 = \min [   y(s_1), y(w_3) = P_{2,M}, y(w_4)].
	\end{gather*}
	In the example of Fig. \ref{fig:PMCregions}, the intersection segment starts at $x_1$ and ends at $v_4$ passing by $i$. Although $\overline{x_1 i} \cup \overline{i v_4}$ is a different segment from $\overline{x_1 v_4}$, their projections over the $(P_1,P_2)$ plane are identical, thus we are only interested in segment ends over both sides $S_1$ and $S_2$ (the projection of $\overline{g_3 g_2} \cup \overline{g_2 j_2}$ is the same as that of the segment $\overline{g_3 j_2}$).\commentaire{ However, for the case of $S_U$, the intersection point of $\mathcal{PL}_2$ and $\mathcal{PL}_4$ has an impact over the segments end points since the projection of the segments is affected as it is discussed next.} \\
	
	\paragraph{Side $S_U$} Unlike for the other sides, none of $P_1$ or $P_2$ is fixed, but $P_2$ can be expressed in terms of $P_1$; therefore, we evaluate the position of the endpoints of the segments on $S_U$ in terms of maximum $P_1$ and minimum $P_1$.
	\\
	When $L_{\lambda} = \mathcal{PL}_2 \cap \mathcal{PL}_4$ does not intercept $S_U$ as it is the case for Fig. \mbox{\ref{fig:PMCregions}} for example, the intersection of $\mathcal{PL}_{2,4}$ with $S_U$ yields a unique segment ($\overline{v_4 v_5}$ in case of Fig. \mbox{\ref{fig:PMCregions}}).  The endpoint corresponding to the minimum value of $P_1$ is due to the intersection of $\mathcal{PL}_{2,4}$ with either $S_2$ or $\mathcal{PL}_3$.
	\begin{gather}
	\min P_1 = \max [  x(\mathcal{PL}_{2,4} \cap \mathcal{PL}_3 \cap S_U), x(\mathcal{PL}_{2,4} \cap S_2 \cap S_U)  ], \nonumber\\
	\min P_1 = \max [ x(s_U), x(w_5) ]. \label{minP1}
	\end{gather} 
	The maximum value of $P_1$ is due to the intersection of $\mathcal{PL}_{2,4}$ with either $\mathcal{PL}_{1}$ or $S_1$.
	\begin{gather}
	\max P_1 = \min [ x(\mathcal{PL}_{2,4} \cap \mathcal{PL}_1 \cap S_U), x( \mathcal{PL}_{2,4} \cap S_1 \cap S_U) ], \nonumber\\
	\max P_1 = \min [ x(x_U), x( w_4) ]. \label{maxP1}
	\end{gather}
	If $i$ resides on $S_U$, then the intersection segment of $\mathcal{PL}_{2,4}$ with $S_U$ is broken into two segments. In that case, if we let $a$ and $b$ be the points given by \mbox{\eqref{minP1}} and \mbox{\eqref{maxP1}} respectively, then the optimization over $S_U$ has to be conducted separately over $\overline{bi}$ from the side of region 1, and over $\overline{ia}$ from the side of region 2. In this case, $i$ corresponds to the $\max P_1$ point in $\overline{ia}$ and to the $\min P_1$ point in $\overline{bi}$. Assuming the conditions of the last row in Table \mbox{\ref{tabu:segments}}, this is the only case where 4 segments in total have to be checked to find the optimal D2D throughput achieving point. The coordinates of all the points mentioned in this section are provided in Appendix \ref{app:EndPoints}.
	\subsection{D2D Throughput Optimization}\label{sec:PowerOptimization}
	At last, given the segments locations and endpoints, the analytical power optimization can be conducted. The mathematical formulation varies according to the side the segment is included in. 
	
	\subsubsection{\underline{\textbf{Side} $\boldsymbol{S_1}$}}\label{Side1} The optimization variable is $P_1$ and the problem formulation is the following: 
	\begin{gather*}
	P_1^* =\mathop{\arg\max}_{P_1} B\log_2(1+\frac{P_1h_d}{P_2\eta_2+\sigma^2}) + B\log_2(1+\frac{P_2h_d}{P_1\eta_1+\sigma^2})
	\end{gather*} 
	\centering s.t. $P_1 \in [\min P_1, \max P_1] \& \quad P_2 = P_{2,M}$\\
	\justify
	Taking the derivative of $F(P_1) = R_{D2D}(P_1,P_{2,M})$  with respect to $P_1$, we get: $$\frac
	{\partial F}{\partial P_1}\frac{\ln 2}{B} =\frac{h_{d}}{P_1h_d+ P_{2,M}\eta_2+\sigma^2} + \frac{-\eta_1 P_{2,M} h_d}{(P_1\eta_1 + \sigma^2)(P_1\eta_1 + P_{2,M} h_d + \sigma^2)}$$
	The sign of $\partial F/\partial P_1$ is equal to the sign of the following second-degree polynomial of $P_1$:
	$$	P_1^2 \underbrace{\eta_1^2}_a+ P_1\underbrace{2\eta_1\sigma^2}_b   +\underbrace{P_{2,M}(h_d - \eta_1) \sigma^2 - P_{2,M}^2 \eta_2\eta_1+ \sigma^4}_c $$
	It is shown in Appendix \ref{app:S_1} that independently of the sign of the polynomial's discriminant, it is sufficient to test which of $\min P_1$ or $\max P_1$ delivers the best throughput and then select the corresponding segment endpoint. The coordinates of the endpoint form the optimal triplet ($P_{1}^*,P_2^*,P_u^*$) maximizing the D2D throughput over the side $S_1$. The endpoint coordinates are given in Appendix \ref{app:EndPoints}.
	\subsubsection{\underline{\textbf{Side} $\boldsymbol{S_2}$}} Following the same reasoning as for $S_1$ (with the only difference that the optimization variable is now $P_2$ instead of $P_1$, and $P_1 = P_{1,M}$), the same conclusion is reached, i.e. the maximum D2D throughput is delivered by the points corresponding either to $\min P_2$ or to $\max P_2$.
	
	\subsubsection{\underline{\textbf{Side} $\boldsymbol{S_U}$}} Since $i$ is accounted for in the maximum and minimum values of $P_1$ for each intersection segment, the optimization can thus be conducted over each segment independently.
	
	The D2D throughput maximization problem over the intersection segment of $\mathcal{PL}_2$ with $S_U$ can be written as follows: 
	\begin{gather*}
	P_{1}^{*}=  \mathop{\arg\max}_{P_1} F(P_1,P_2)
	\end{gather*}
	\centerline{s.t $P_{u,M}  =\frac{ P_1 \eta_1 + P_2 h_{d}}{h_{d_1,u}}, P_1 \in \mathbb{U} = [\min P_1, \max P_1], P_u = P_{u,M}.$}
	Replacing $P_2$ by $({P_{u,M} h_{d_1,u} - P_1 \eta_1})/{h_{d}}$ in $F(P_1,P_2) = B \log_2 ( 1 + \frac{P_1 h_d}{P_2 \eta_2 + \sigma^2}) + B\log_2 ( 1 + \frac{P_2 h_d}{P_1 \eta_1 + \sigma^2})$,  we get: 
	\begin{gather*}
	F(P_1) =B \log_2(1+\frac{P_1h_d^2}{({P_{u,M} h_{d_1,u} - P_1 \eta_1})\eta_2+{h_{d}}\sigma^2}) 
	\\+
	B \log_2(1+\frac{P_{u,M} h_{d_1,u} - P_1 \eta_1}{P_1\eta_1+\sigma^2}).
	\end{gather*}
	Since $P_2>0$, we must have $P_1 < P_{u,M}h_{d_1,u}/\eta_1$ (which adds to the constraint of $\max P_1$). Taking  the derivative of $F$ with respect to $P_1$ leads to:  $\ln(2)\partial F/ \partial P_1 B$=
	\begin{gather*}
	= 
	\frac{ h_{d}^2 ( h_{d_1,u}\eta_2P_{u,M} +\sigma^2h_d)/[P_{u,M} h_{d_1,u}\eta_2 - P_1 \eta_1\eta_2+{h_{d}}\sigma^2] }
	{
		(P_{u,M} h_{d_1,u}\eta_2 - P_1 (\eta_1\eta_2 - h_d^2)
		+h_d\sigma^2) 
	}
	\\
	-
	\frac{ \eta_1}
	{
		(P_1\eta_1+\sigma^2)
	}.
	\end{gather*}
	
	Since $P_1 \in \mathbb{U}$, it can be easily verified that both denominators are positive; therefore, only the numerator is needed to evaluate the sign of $\partial F / \partial P_1$:
	\begin{align*}
	\sign \frac{\partial F}{\partial P_1} = \sign [ 
	& h_{d}^2 ( h_{d_1,u}\eta_2P_{u,M} +\sigma^2h_d)
	(P_1\eta_1+\sigma^2)  
	\\
	&-\eta_1
	(P_{u,M} h_{d_1,u}\eta_2 - P_1 \eta_1\eta_2+{h_{d}}\sigma^2)\\
	&
	(P_{u,M} h_{d_1,u}\eta_2 - P_1 (\eta_1\eta_2- h_d^2) +{h_{d}}\sigma^2)].
	\end{align*} After some simplifications and re-arrangements, the sign of $\partial F/ \partial P_1$ can be written as the sign of a second-degree polynomial of $P_1$ of the form $AP_1^2 + BP_1 + C$ with: 
	\begin{gather*}
	A = -(\eta_1\eta_2-h_d^2)\eta_1^2\eta_2; \quad 
	B = 2\eta_1^2\eta_2 ( P_{u,M} h_{d_1,u}\eta_2 + \sigma^2h_d);\\
	C = -P_{u,M}^2 h_{d_1,u}^2 \eta_2^2 \eta_1 + P_{u,M}\sigma^2 h_dh_{d_1,u}\eta_2 (h_d - 2\eta_1) \\+ \sigma^4h_d^2(h_d-\eta_1).
	\end{gather*}
	Given the root $sol_1 = (-B-\sqrt{ B^2 - 4 AC})/2A$ of the polynomial,  
	we show in Appendix \ref{app:S_u} that $P_1^*$ is either given by
	$\min P_1$, $\max P_1$, or $sol_1$ (when it is included in the interval $[\min P_1,\max P_1]$), according to the value delivering the highest throughput. Regarding the optimization over the intersection segment of $\mathcal{PL}_4$ with $S_U$, the same steps are followed to determine the optimal value of $P_1$: we start by writing the expression of $F(P_1)$ by replacing $P_2$ in $F(P_1,P_2)$ with $(P_{u,M} h_{d_2,u}-P_1 h_d) / \eta_2)$. Then, the study of the sign of $\partial F / \partial P_1$ turns into the study of the sign of another second-degree polynomial $A^{'}P_1^2 + B^{'}P_1 + C^{'}$ with: 
	\begin{gather*}
	A^{'} = (\eta_1\eta_2-h_d^2)\eta_1; \quad 
	B ^{'}= 2\eta_1 ( P_{u,M} h_{d_2,u}h_d + \sigma^2\eta_2); \nonumber\\
	C^{'} = -P_{u,M}^2 h_{d_2,u}^2\eta_1 - \sigma^2\eta_1h_{d_2,u}P_{u,M}  
	+ \sigma^4(\eta_2 - h_d).
	\end{gather*}
	Also, following the different channel conditions concerning $\sign (\eta_1\eta_2 - h_d^2)$, and considering all the possible relative positions between \commentaire{$sol^{'}_2,$}$\max P_1,\min P_1$, and $sol_1^{'}$, the same result as previously is obtained,  which can be cast as: $$P_{1}^{*} = \arg\max [  F(\min P_1), F(\max P_1), F(sol^{'}_1)].$$
	As a conclusion, the optimization over the sides $S_1$ and $S_2$ resides in selecting the corresponding endpoint achieving the highest throughput. On the side $S_U$, a maximum of three additional points ($i,sol_{1},sol_{1}^{'}$) may need to be considered to get the highest D2D throughput.
	\subsection{Summary of the PA Procedure and Extension to the Second Decoding Order}
	In this section, the geometrical representation of the FD-SIC PA problem allowing for a drastic reduction of the search space size was described. It was shown that the initial search volume in Sec. \ref{sec:SolutionRepresentation} can be reduced to a set of intersection segments (Sec. \ref{sec:SpaceReduction}) from which a subset is selected (\ref{sec:UsefulIntersections}). These segments search spaces are then further reduced to become a finite set of points (Secs. \ref{sec:SegmentEndpoints}, \ref{sec:PowerOptimization}). In the worst case scenario, the original PA problem, which had $2^{12}-1$ variants, is converted into the search for the maximum throughput of a list of seven elements: two elements from $S_1$, two from $S_2$ and three additional elements from $S_U$ ($w_4$ is a common endpoint to $S_1$ and $S_U$, and $w_5$ is common to $S_2$ and $S_U$). The global PA procedure to determine the optimal D2D rate for the first decoding order of FD-SIC is summarized in algorithm \ref{algo:GlobalPA}. 
	
	\begin{algorithm}[h]
		\SetAlgoLined
		\SetKwInOut{Input}{input}\SetKwInOut{Output}{output}
		\Input{$P_{1,M}$, $P_{2,M}$, $P_{u,M}$, $P_{u,m}$, $h_{b,u}$, $h_d$,  $\eta_1$, $\eta_2$, $h_{d1,u}$, $h_{d_2,u}$, $h_{d_1,b}$, $h_{d_2,b}$}
		\KwResult{Optimal triplet $(P_1^{*}, P_2^{*},P_u^{*})$.}
		\caption{Optimal PA procedure for FD-SIC}
		\eIf 
		{\eqref{equ:sufficientPMC2} $\wedge$ \eqref{equ:sufficientPMC4} $\wedge$ \eqref{equ:Pmax}}
		{
			Test $\Omega,\xi,\gamma,\tau$ and build $PMC_{2,4}$\;
			Execute Algorithm \ref{alg:SideSelection} to determine $x_i$ and $s_j$\;
			Follow Table \ref{tabu:segments} to keep the necessary segments\;
			Compute $R_{D2D}$ for the edges of each segment\;
			Keep the point providing the highest throughput.
			
		}
		{Empty search space, no solution}
		\label{algo:GlobalPA}
	\end{algorithm}
	Regarding the resolution for the second decoding order, the PA procedure itself is unchanged, but the changes in $PMC_1$ and $PMC_3$ lead to some modifications.  Here is the list: 
	\begin{itemize}
		\item Modification in the expressions of $PMC_1$ and $PMC_3$:
		\begin{align*}
		&P_{u}h_{b,u} < P_{1}h_{b,d_1} - P_2 h_{b,d_2}    & (PMC_1)\\
		&P_u h_{b,u} < P_2 h_{b,d_2}  & (PMC_3)\\
		\end{align*}
		\item The necessary and sufficient conditions \eqref{equ:sufficientPMC2}, \eqref{equ:sufficientPMC4} and \eqref{equ:Pmax} become:
		\begin{gather}
		h_{d_1,u}h_{b,d_2} - h_d h_{b,u} >  2\frac{ \eta_1 h_{b,u}  h_{b,d_2}}{h_{b,d_1}} \tag{\ref{equ:sufficientPMC2}}
		\\
		h_{d_2,u}h_{b,d_2} - \eta_2 h_{b,u} >  2\frac{ h_{b,u} h_d h_{b,d_2}}{h_{b,d_1}} \tag{\ref{equ:sufficientPMC4}}
		\\
		2P_{u,m}\frac{h_{b,u}}{ h_{b,d_1}} < P_{1,M} \quad \&\& \quad  
		P_{u,m}\frac{h_{b,u}}{h_{b,d_2}} < P_{2,M} \tag{\ref{equ:Pmax}}
		\end{gather}
		\item Concerning section \ref{sec:SegmentEndpoints}, the roles of $PMC_1$ and $PMC_3$ are interchanged concerning the settlement of the segment endpoints.  
		\item The three non-occuring ($x_i,s_i$) pairs of section \ref{sec:SideSelection} become: $(x_1,s_2), (x_1,s_U), (x_U,s_2)$.
	\end{itemize} 
	Sections \ref{sec:PMC24} and \ref{sec:PowerOptimization} are kept unchanged because building $PMC_{2,4}$ is independent of $PMC_1$ and $PMC_3$, and given the endpoints of the segments subset, the optimization of section \ref{sec:PowerOptimization} is not affected by the change in $PMC_1$ and $PMC_3$.

	\section{Channel Allocation}\label{Munkures}
	In this section, the procedure for optimal channel allocation to D2D devices is conducted. Recalling that the D2D system is underlaying a pre-established CU network, D2D channel allocation is equivalently referred to as D2D-CU pairing.
	
	Having determined the analytical PA solutions for all the transmission scenarios, their resolution cost is a constant-time operation. Therefore, filling the D2D rate tables $\mathcal{R}_{D2D}^{FD-NoSIC},\mathcal{R}_{D2D}^{HD-NoSIC},\mathcal{R}_{D2D}^{HD-SIC}$, and $\mathcal{R}_{D2D}^{FD-SIC}$ for every D2D-CU pair is accomplished with a complexity in $O(KD)$. In the case of FD-SIC, the channel links, required CU rate and transmit power limits of a D2D $n$ and a CU $u_i$ may be such that one of the conditions \eqref{equ:sufficientPMC2}, \eqref{equ:sufficientPMC4}, \eqref{equ:Pmax}  is not valid. If this is the case for both decoding orders, then the PA of FD-SIC reverts to that of FD-NoSIC to fill the element $\mathcal{R}_{D2D}^{FD-SIC}(n,i)$ as explained in the end of section \ref{FD-SIC}. Also, if both decoding orders are possible for this combination, $\mathcal{R}_{D2D}^{FD-SIC}(n,i)$ is filled with the highest rate among the two possible orders. When filling matrix $\mathcal{R}_{D2D}^{HD-SIC}$, and as explained in section \ref{HD-SIC}, HD-SIC reverts to HD-NoSIC in any of the two half-time slots, when conditions \eqref{necessary1} or \eqref{necessary2} are not valid. Given these rate tables, the optimal channel allocation tables $O_{FD-NoSIC}^{*}, O_{HD-NoSIC}^{*}, O_{FD-SIC}^{*}$, and $O_{HD-SIC}^{*}$ corresponding to every transmission scenario are obtained by solving the channel assignment problem in a way to maximize the total D2D throughput. This problem takes the generic formulation given by:
	\begin{gather*}
	O^{*} = \mathop{\arg\max}_{(i,j)\in \Iintv*{1,D}\times  \Iintv*{1,K}} (\sum_{i=1}^{D}\sum_{j=1}^{K}\mathcal{R}_{D2D}(i,j)\times o(i,j))\\
	\text{s.t. the constraints of \eqref{MunkresConditions} are verified}   
	\end{gather*} 
	This assignment problem is efficiently solved by the Kuhn-Munkres (KM) algorithm \mbox{\cite{Munkres}}, also called the Hungarian method, with a complexity of $O(D^2K)$ \mbox{\cite{cui2016solving}}. Note that the global resource allocation complexity is now dominated by that of the channel assignment after the important PA complexity reduction. The KM can be directly applied in our study to yield the optimal channel assignment by rewriting the problem as a minimization of the opposite objective function ($- \mathcal{R}_{D2D}$). As a conclusion, the optimal PA procedures allowed for an efficient filling of the rate tables which are then fed to the KM solver. This delivers the global optimal solution of the joint channel and power allocation problem formulated in section \ref{ProbFormulation}.
	

	\section{Numerical Results}\label{numResults}
	In our simulation setup, the BS is positioned at the center of a hexagonal cell with an outermost radius of $300$~m. The D2D users and the CU users are randomly located within the cell. The distance between the D2D users of every pair is below a maximum value $d_{max}$. The propagation model includes large-scale fading with a path-loss exponent $\alpha=3.76$, and an $8$~dB zero mean lognormal shadowing. The maximum transmit power of the devices and CU is $24$~dBm. The system bandwidth is $20$~MHz, divided into $N=64$ channels, leading to a UL bandwidth of $B=312.5$~kHz, with a noise power of $-119$~dBm. The minimum required rate $R_{u,min}$ is the same for all the CU users, and the SI cancellation factor $\eta$ is the same for all D2D pairs, its value being varied between $-130$ and $-80$ dB. The results are averaged over 1000 different realizations of the devices and CU positions. Unless specified otherwise, $R_{u,min}$ is set to $1.5$~Mbps, $K$ is set to 20 CUs, $d_{max}=100$~m, and $D = 5$ D2D pairs.
	\begin{figure}[h]
		\centering
		\begin{subfigure}{0.49\linewidth}
			\includegraphics[width=\textwidth]{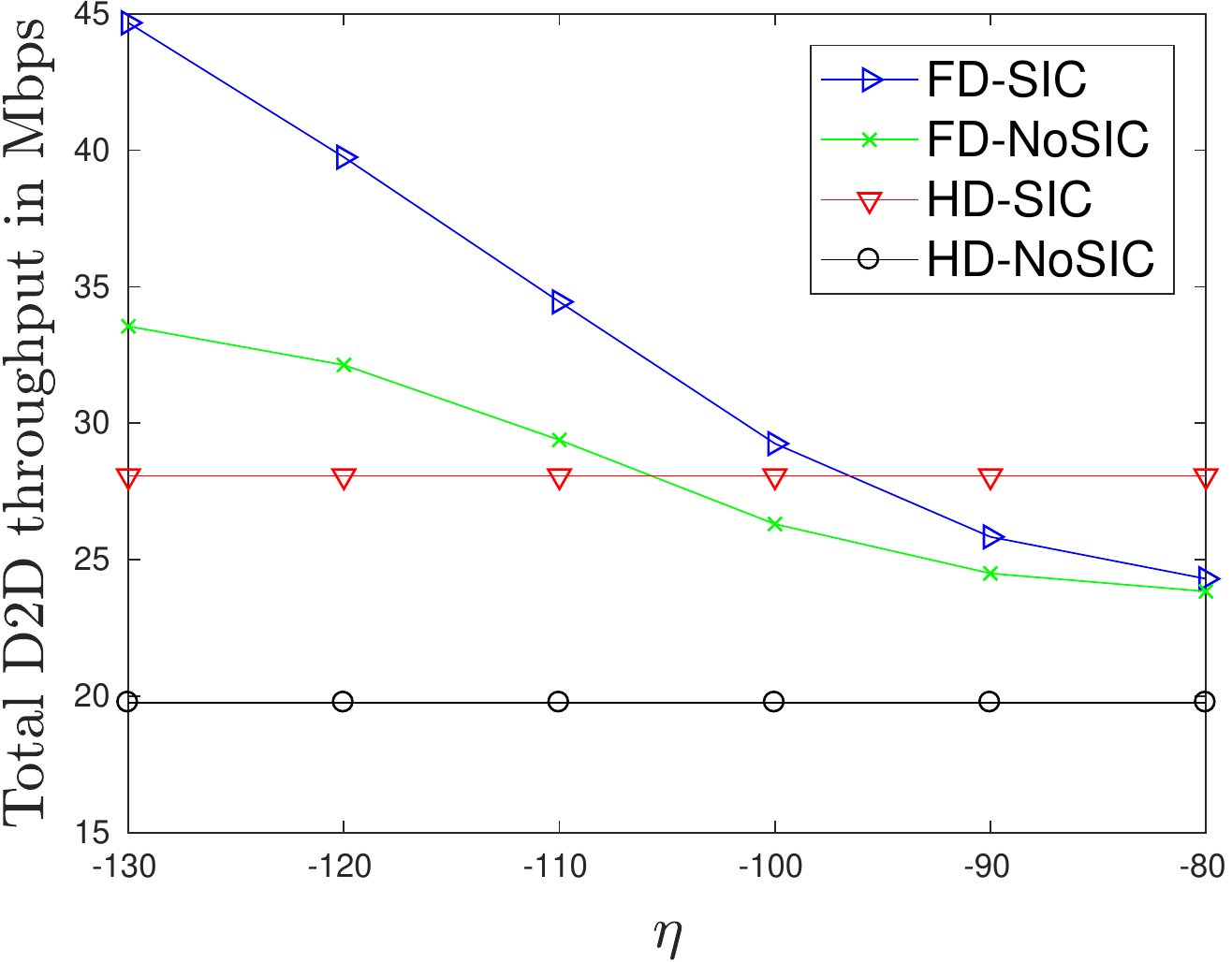}
			\caption{$R_{u,min}=1.5$~Mbps}\label{figResult1a}
		\end{subfigure}
		\begin{subfigure}{0.49\linewidth}
			\includegraphics[width=\textwidth]{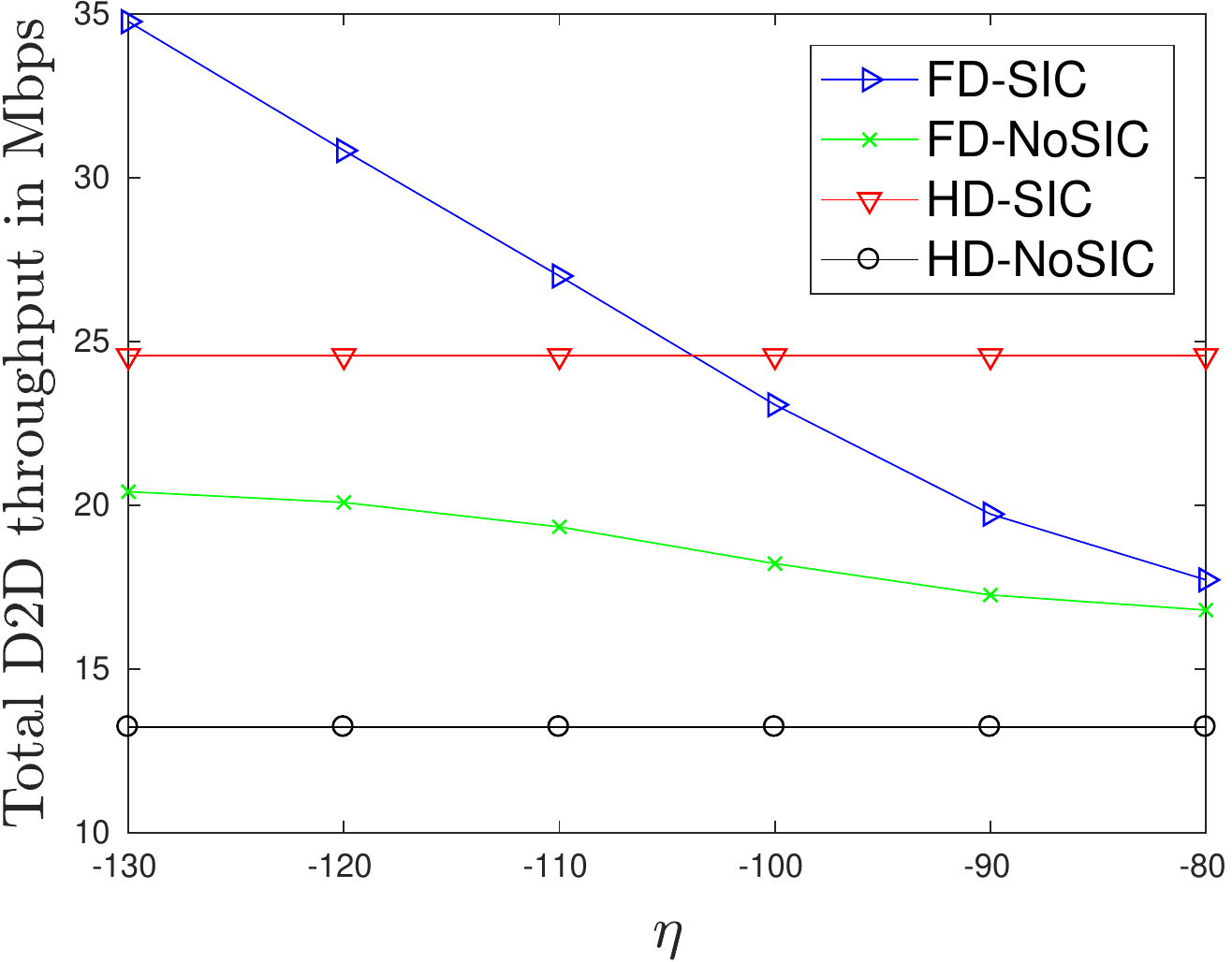}
			\caption{$R_{u,min}=3$~Mbps}\label{figResult1b}
		\end{subfigure}
		\caption{Total D2D throughput as a function of $\eta$ for $K=20$ CUs, $D=5$ D2D pairs, and $d_{max}=100$~m.}
		\label{figResult1}
	\end{figure}
	Figure \ref{figResult1} presents the total D2D throughput as a function of $\eta$, for two different values of $R_{u,min}$. At first, it can be noted that HD schemes are not affected by $\eta$ unlike FD schemes. This was expected since self-interference occurs only for FD transmission. Secondly, the mutual SIC enabled schemes outperform their counterpart No-SIC schemes for both HD and FD transmission scenarios. Indeed, a 41~\% rate increase is observed in Fig. \ref{figResult1a} between HD-SIC and HD-NoSIC (going from 19.8~Mbps to 28.1~Mbps). The throughput enhancements due to mutual SIC for the case of FD transmission vary between a 2~\% increase for $\eta = -80$~dB, to  $33$~\% increase for $\eta=-130$~dB. The performance gains of FD-SIC with respect to FD-NoSIC increase with the SI cancellation capabilities of the devices because of two reasons: on the one hand, the decrease of $\eta$ relaxes the constraints \eqref{necessary3} and \eqref{necessary4}, thereby increasing the number of D2D-CU pairs that benefit from FD-SIC (from an average of $0.36$ FD-SIC D2D pairs for $\eta = -80$~dB to $1.92$ pairs for $\eta=-130$~dB, with $R_{u,min} = 1.5$ Mbps). On the other hand, the decrease of $\eta$ reduces the interference terms in the D2D throughput expression, which translates into a higher achieved throughput. 
	
	As expected, when comparing the performance for different required CU rates between Figs. \ref{figResult1a} and \ref{figResult1b}, the increase of $R_u$ from $1.5$ Mbps to $3$ Mbps decreases the achieved D2D throughput for all  proposed methods. However, the percentage gain in the performance of SIC procedures with respect to NoSIC increases from 41~\% to 86~\% for the HD case, and from 33~\% to 70~\% for the FD case (for $\eta = -130$~dB). The reason behind this gain increase is that NoSIC algorithms are highly affected by the value of $P_u$ ($ \geq P_{u,m}$) since they suffer from its interference, which is not the case of SIC techniques. In fact, even though the total number of FD-SIC enabled D2D-CU pairs decreases with $R_{u,min}$ (due to harsher mutual SIC constraints, cf. eq. \eqref{equ:Pmax}), the  Munkres allocation yields an increasing number of selected D2D-CU pairs achieving FD-SIC (or HD-SIC) with $R_{u,min}$ (from an average of 0.8 for $R_{u,min} = 1.5$~Mbps to an average of 1.24 for $R_{u,min}=3$~Mbps, with $\eta = -90$~dB). This corroborates the idea that the throughput decrease of No-SIC techniques with $R_{u,min}$ is more important than that of SIC techniques, to a point where the contribution of mutual SIC techniques in maximizing the throughput is more prominent when $R_{u,min}$ increases. This is verified by comparing the percentage decrease of D2D throughput for every algorithm when moving from $R_u =1.5$~Mbps to $R_u=3$ Mbps: a decrease of 39~\%, 33~\%, 22~\%, and 13~\% is observed for the algorithms FD-NoSIC, HD-NoSIC, FD-SIC, HD-SIC respectively. The greater decrease of FD-NoSIC performance compared to HD-NoSIC justifies the shift of the intersection point between FD-SIC and HD-SIC to the left when $R_{u,min}$ increases. Indeed, as explained in Section \ref{Munkures}, FD-SIC and HD-SIC are applied when possible, on top of FD-NoSIC and HD-NoSIC respectively. If the performance gap between FD-NoSIC and HD-NoSIC diminishes, HD-SIC will outperform FD-SIC over a broader span of $\eta$ values before FD-SIC eventually catches up and surpasses HD-SIC for smaller $\eta$ values (i.e. for better SI cancellation capabilities of the devices).	
	\begin{figure}[h]
		\centering
		\includegraphics[width=0.75\textwidth/2]{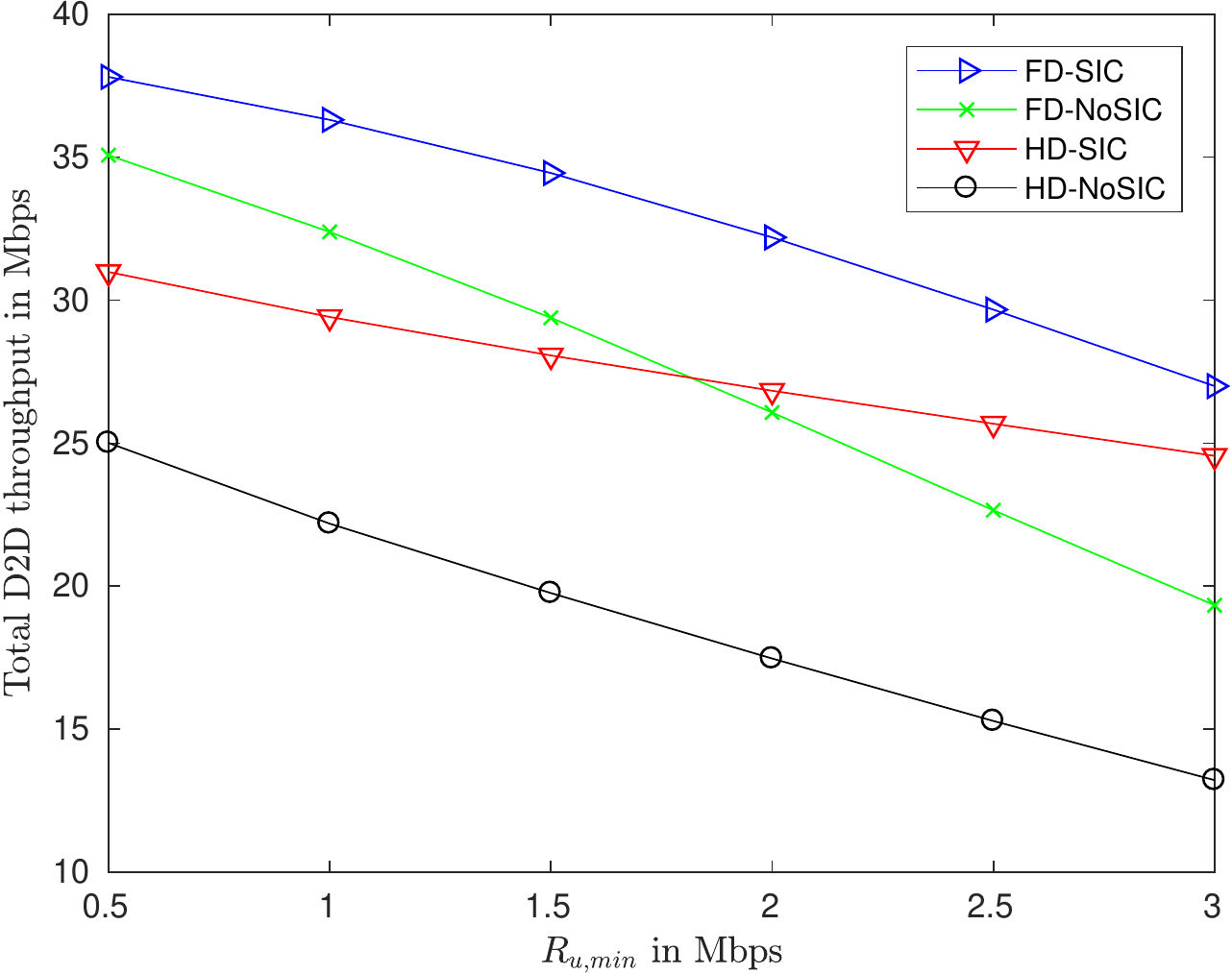}
		\caption{Total D2D throughput as a function of $R_{u,min}$ for $\eta=-110$~dB.}
		\label{figResult2}
	\end{figure}
	This evolution of FD-SIC and HD-SIC can also be observed from another perspective in Fig. \ref{figResult2}, where the total D2D throughput is presented as a function of the CU required rate. In the conditions of Fig. \ref{figResult2}, the gap between FD-NoSIC and HD-NoSIC is large enough so that no intersection occurs between FD-SIC and HD-SIC. However, it can still be observed that the gap between FD-SIC and HD-SIC reduces as the CU required rate increases. 
	
\begin{figure}[h]
		\centering
		\includegraphics[width=0.75\textwidth/2]{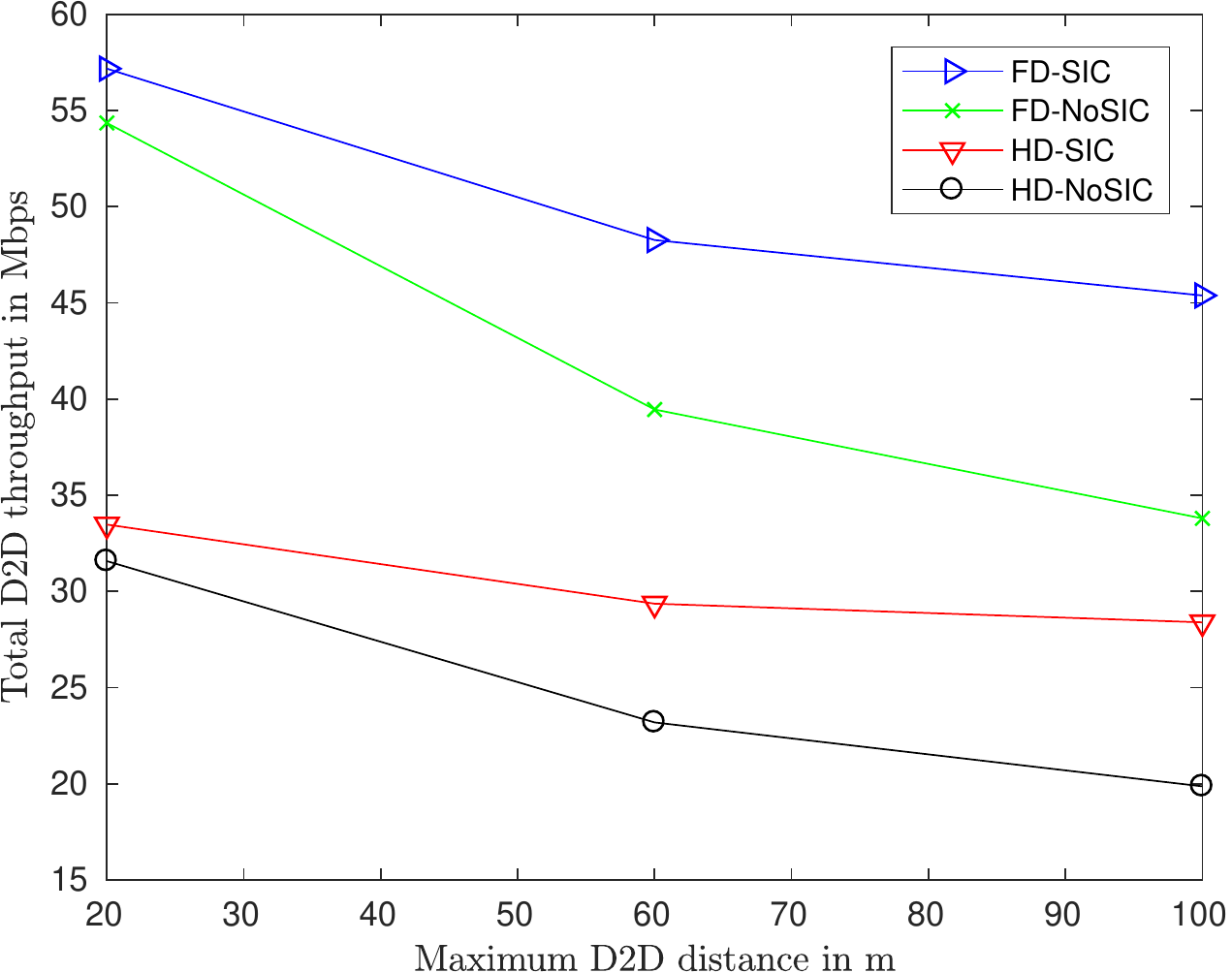}
		\caption{Total D2D throughput as a function of $d_{max}$ for $\eta=-130$~dB.}
		\label{figResult3}
	\end{figure}

	In Fig. \ref{figResult3}, the variation of the total D2D throughput is presented as a function of the D2D maximum user distance $d_{max}$. The increase of $d_{max}$ leads to a significant decrease in the performance of all proposed methods since $h_d$, the channel gain of the direct link between $d_1$ and $d_2$, is reduced on average. However, this increase of $d_{max}$ is accompanied by a greater percentage increase in performance due to mutual SIC for FD and HD transmission scenarios, with respect to No-SIC scenarios. Indeed, FD-SIC achieves a D2D throughput 128~\% higher than FD-NoSIC for $d_{max}=100$~m, compared to the 81~\% increase achieved for $d_{max}=20$~m. This is due to having more FD-SIC enabled D2D-CU pairs when distancing the D2D users further apart from one another, since an average of 1.96 pairs successfully apply FD-SIC for $d_{max}=20$~m as opposed to 3.33 pairs for $d_{max} = 100$~m. The reason behind this increase is the decrease in $h_d$ which relaxes the sufficient conditions \eqref{equ:sufficientPMC2} and \eqref{equ:sufficientPMC4}, thereby enabling more FD-SIC cases.
	
\begin{figure}[h]
		\centering
		\includegraphics[width=0.75\textwidth/2]{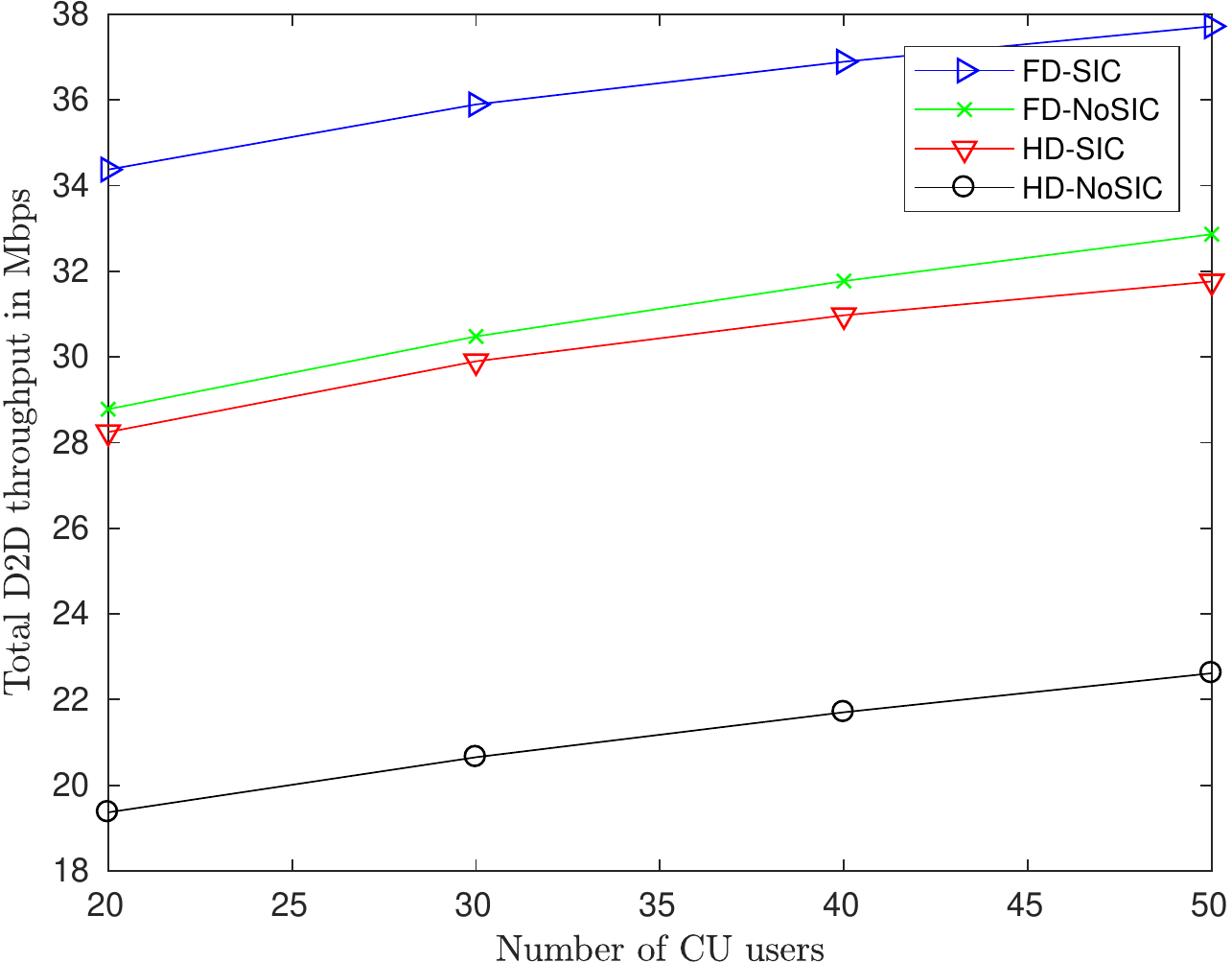}
		\caption{Total D2D throughput as a function of $K$ for $\eta=-110$~dB.}
		\label{figResult4}
	\end{figure}

\begin{figure}[h]
		\centering
		\begin{subfigure}{0.49\linewidth}
			\includegraphics[width=\textwidth]{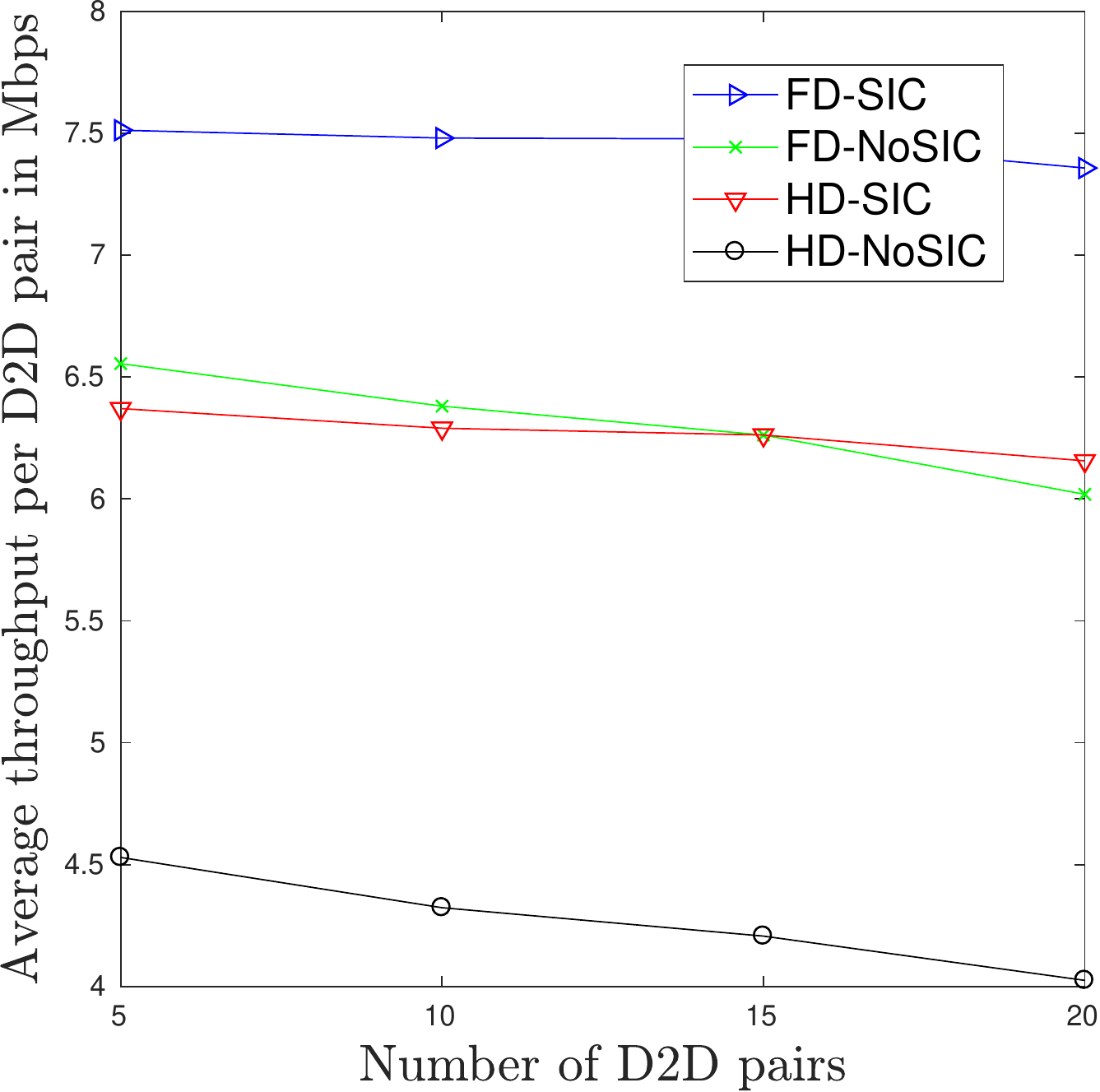}
			\caption{Average D2D throughput}\label{figResult5a}
		\end{subfigure}
		\begin{subfigure}{0.49\linewidth}
			\includegraphics[width=\textwidth]{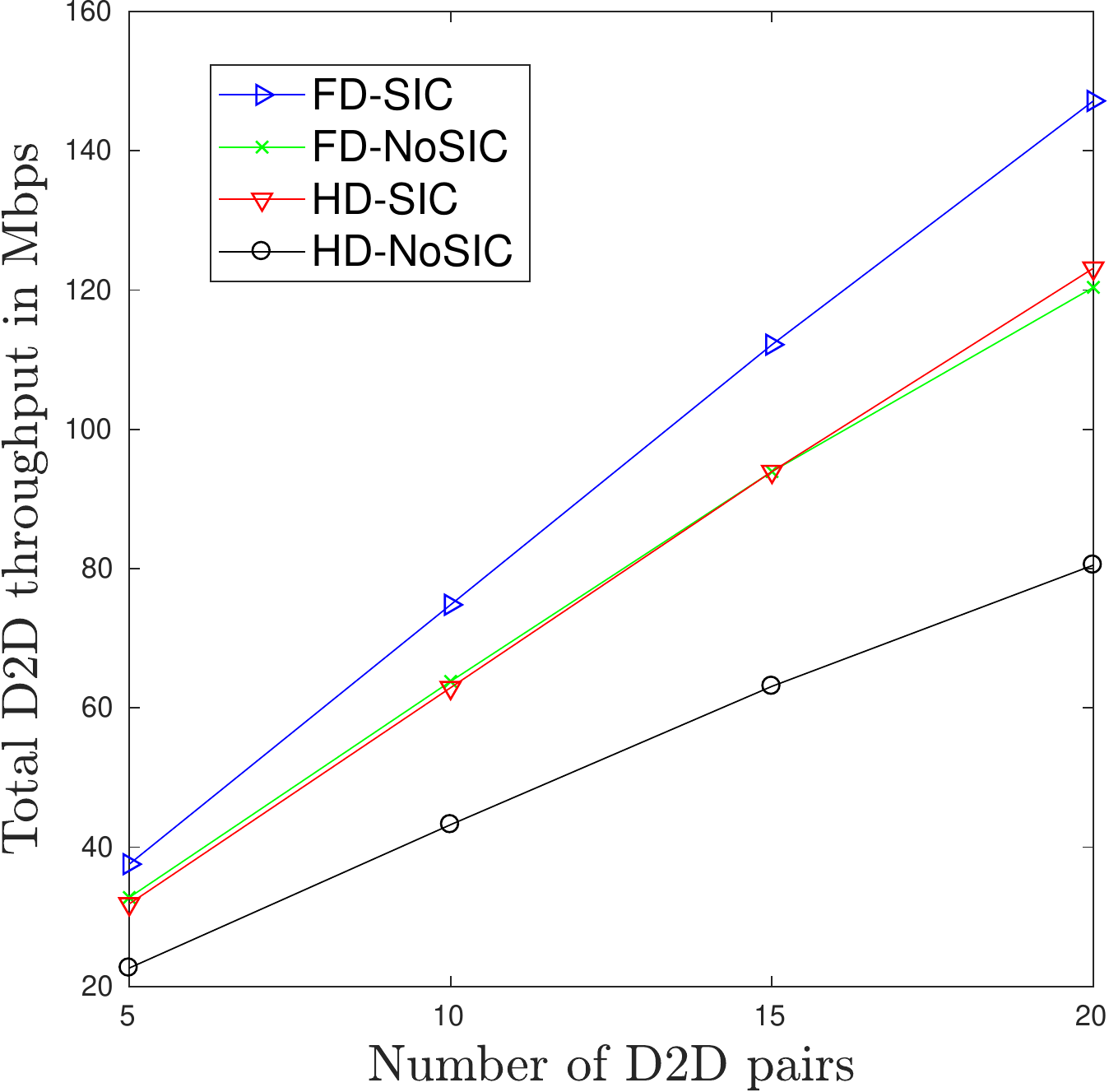}
			\caption{Total D2D throughput}\label{figResult5b}
		\end{subfigure}
		\caption{Total and average D2D throughput as a function of the number of D2D pairs for $K=50$ CUs and $\eta=-110$~dB. }
		\label{figResult5}
	\end{figure}

	Fig. \ref{figResult4} presents the evolution of the D2D throughput as a function of the number of CU users in the cell. Although the channel properties of the D2D users (i.e. $h_d$, $h_{d_1,u}$ and $h_{d_2,u}$) are unchanged, the total D2D throughput of all  techniques benefits from the additional diversity provided by the greater number of CU users. This also favors the FD-SIC enabled pairs, as their average number grows from 1.98 for $K=20$ to 2.46 for $K=50$. We can therefore conclude that the important performance gain achieved by SIC methods, with respect to No-SIC methods, can be obtained without requiring the implementation of SIC at all D2D and CU receivers. Indeed, generally only 2 or 3 triplets need to perform SIC which is enough to boost the D2D system capacity, while the others can settle for the simple classical No-SIC receivers. Therefore, the additional complexity is localised at the level of the users performing SIC for which the major throughput increase is worth the incurred SIC complexity.
	Finally, the total and average throughput variations are presented in Fig. \ref{figResult5} as a function of the number of D2D pairs in the system, for a fixed value of $K=50$. In Fig. \ref{figResult5a}, the average throughput per D2D pair is shown to slightly decrease  with the increasing number of D2D pairs. In a sense, this is the dual of the behavior observed in Fig. \ref{figResult4}, since the ratio $K/D$ decreases with $D$ and thus the system diversity - in terms of the average number of possible CU channel choices for every D2D pair to be collocated on - decreases, thus reducing the achievable throughput per D2D pair. Nonetheless, the total throughput follows a quasi linear progression with the number of D2D pairs because the additional D2D pairs are allocated on orthogonal channels, therefore each D2D pair can be associated more or less to an additional D2D rate unit.  
	Figs. \ref{figResult4} and \ref{figResult5} indicate that, for a fixed number of $D2D$ users or CU users, the effect of the proportion $K/D$ on the average D2D throughput per D2D pair is rather limited. The most dominant factors remain the distance between D2D users, the SI cancellation capabilities of the receivers (for FD-SIC), and the required CU rate.
	
	\section{Conclusion}\label{sec:conclusion}
	In this paper, the use of NOMA with mutual SIC was proposed for the first time between cellular users and FD-D2D devices underlaying the cellular channels. The necessary and sufficient conditions for applying FD-SIC were derived and a highly efficient PA procedure was elaborated \commentaire{introduced}to solve, in constant time operation, the throughput maximization problem of significant original complexity. The optimal, yet simple, PA resolution allowed for achieving global optimal resource allocation by conveniently combining the Kuhn-Munkres channel assignment with the proposed PA methods.
	The results show important performance gains obtained by applying SIC in D2D underlay systems in both HD and FD transmission schemes, promoting thereby the use of mutual SIC NOMA for D2D systems whenever possible. When applying mutual SIC, the comparison between HD and FD transmission scenarios showed that FD-SIC is  more efficient  for average to high SI cancellation capabilities, moderate CU rate requirements and significant D2D distances, while HD-SIC performs better especially at low SI cancellation capabilities. Future work directions of the study could be to adapt the PA procedure for D2D underlay to downlink cellular systems which present different sets of challenges and networking paradigms. Moreover, the integration of our work in UAV-assisted NOMA networks for IoT could also be considered, as in the context of \mbox{\cite{liu2019dsf}}.

	\section*{Acknowledgment}
	This work has been funded with support from IMT Atlantique and the Lebanese University Research Support Program.\\
	
	
	\appendices
	\section{} \label{app1}
	To determine if the search space is totally included in region 1 or 2 for the cases 3) and 4), we introduce $f_1, f_3$ and $f_{\lambda}$, the functions of $P_1,P_2$ which yield the $P_u$ value corresponding to the planes $\mathcal{PL}_1, \mathcal{PL}_3$ and to $L_{\lambda}$. 
	A parametric equation of $L_{\lambda}$ is given by:
	\begin{gather*}
	L_{\lambda} =\begin{cases*}
	x = (\frac{h_d}{h_{d_1,u}}  - \frac{\eta_2}{h_{d_2,u}})m = (\tan  (\gamma) - \tan (\tau))m\\
	y = (\frac{h_d}{h_{d_2,u}}  - \frac{\eta_1}{h_{d_1,u}})m = (\tan (\xi) - \tan (\Omega))m\\
	z = \frac{h_d^2 - \eta_1\eta_2}{h_{d_1,u}h_{d_2,u}} m
	\end{cases*}
	\end{gather*}
	
	In the case of Fig. \ref{fig:PMC2toPMC4}, the search space is included in region 2 if and only if $L_{\lambda}$ is on top of $\mathcal{PL}_1$. For the case of Fig. \ref{fig:PMC4toPMC2}, the search space is included in region 1 if and only if $L_{\lambda}$ is on top of $\mathcal{PL}_3$. To determine the conditions of each scenario, we first have to check if the conditions of case 3), where $\gamma  > \tau $ and $ \xi > \Omega$, or  those of case 4), where $\gamma  < \tau$ and $ \xi < \Omega$, are met. 
	To study the relative position of $L_{\lambda}$ with respect to $\mathcal{PL}_1$ and $\mathcal{PL}_3$, $m$ is chosen such that the comparison is conducted in the first octant. Since in case 3), $\gamma > \tau \Rightarrow \tan(\gamma) - \tan (\tau) > 0$, then $m$ must be positive in case 3), and conversely,  negative in case 4).\\
	The search space is included in region 2 if: 
	\begin{gather*}
	f_{\lambda} (P_1,P_2) > f_1(P_1,P_2) \\
	\Rightarrow	\frac{h_d^2 - \eta_1\eta_2}{h_{d_1,u}h_{d_2,u}} m > \frac{P_2 h_{b,d_2} - P_1 h_{b,d_1}}{h_{b,u}}
	\end{gather*}
	Replacing $P_1$ by $(h_d/ h_{d_1,u} - \eta_2 /h_{d_2,u}) m$, and $P_2$ by $(h_d /h_{d_2,u} - \eta_1 / h_{d_1,u}) m$, we get: 
	\begin{gather*}
	\frac{h_{b,u}(h_d^2 - \eta_1\eta_2)}{h_{d_1,u}h_{d_2,u}} m > 
	(\frac{h_d}{h_{d_2,u}} - \frac{\eta_1}{h_{d_1,u}}) m h_{b,d_2} 
	- (\frac{h_d}{h_{d_1,u}}  - \frac{\eta_2}{h_{d_2,u}} ) mh_{b,d_1}\label{nocooccur}
	\end{gather*}
	Let $\Gamma$ be the proposition \begin{align*}
	\frac{h_{b,u} (h_d^2 - \eta_1\eta_2)}{h_{d_1,u} h_{d_2,u}} > \frac{h_{b,d_2}h_d + h_{b,d_1} \eta_2}{h_{d_2,u}} 
	- \frac{h_{b,d_2}\eta_1 + h_{b,d_1}h_d}{h_{d_1,u}}
	\end{align*}
	Then, we conclude that:
	\begin{itemize}
		\item case 3) ``Search Space included in region 2'' $\Leftrightarrow \Gamma = 1$.
		\item case 4) ``Search Space included in region 2'' $\Leftrightarrow \Gamma = 0$
	\end{itemize}
	On the other hand, the search space is included in region 1 if:
	\begin{gather*}
	f_{\lambda} ( P_1,P_2) > f_3 (P_1,P_2)
	\\
	\Rightarrow	
	(h_d^2 - \eta_1\eta_2) m > \frac{(h_d h_{d_2,u} - \eta_2 h_{d_1,u}) h_{b,d_1}}{h_{b,u}}m
	\end{gather*}
	Let $\Xi$ be the proposition: 
	\begin{equation}
	(h_d^2 - \eta_1\eta_2)h_{b,u} > (h_d h_{d_2,u} - \eta_2 h_{d_1,u})h_{b,d_1}
	\end{equation}
	Therefore, the search space is included in region 1 if $\Xi = 1$ for case 3), and $\Xi = 0$ for case 4). \\
	Conclusion: to determine if  the search space is completely included in one of the two regions, for  case 3) and 4), we simply have to test the validity of $\Gamma$ and $\Xi$ and draw the corresponding conclusion to each case.
	\commentaire{\textcolor{gray}{J'aurais voulu montrer que pour le cas 3) par exemple, $\Gamma$ et $\Theta$ ne peuvent pas etre vraies en meme temps car cela irait a l'encontre d'au moins une des conditions suffisantes. La demonstration est tres simple, j'en suis sur, mais je butte dessus depuis hier. J'y reviendrai plus tard.}
		
		\textcolor{red}{Au pire, si tu en as vraiment besoin, tu peux juste rajouter la phrase: It can be proved that propositions Eta and Teta are exclusive, i.e. cannot be met simulteanously. Car l'article est deja super long... Si ce n'est pas indispensable, il vaut mieux ne rien dire.}
	}
	
	\section{}\label{appC}
	\subsection{Optimal Throughput point over the Side $S_1$}\label{app:S_1}
	The sign of $\partial F/\partial P_1$ is equal to the sign of the following second-degree polynomial of $P_1$: $a P_1^2+ b P_1+ c $, with $a=\eta_1^2, b = 2\eta_1 \sigma^2$, and $c = P_{2,M}(h_d - \eta_1) \sigma^2 - P_{2,M}^2 \eta_2\eta_1+ \sigma^4$. If $\Delta = b^2 - 4ac <0$, the second-degree polynomial is positive, hence the throughput is increasing with $P_1$, and $P_1^*$ is obtained by setting $P_1$ to ${\max P_1}$.\\
	If $\Delta>0$, the polynomial is negative inside the solutions interval, and positive elsewhere. The solutions are: $sol_1 = (-b-\sqrt{\Delta})/2a, sol_2 = (-b+\sqrt{\Delta})/2a.$ Therefore, the throughput is decreasing between $sol_1$ and $sol_2$, then increasing for $P_1 > sol_2$. Since $sol_1 < 0$, three cases are identified depending on the location of $sol_2$ with respect to $\min P_1$ and $\max P_1$: 
	\begin{itemize}
		\item $sol_2 < \min P_1$: the throughput increases with $P_1 \Rightarrow P_{1}^* = \max P_1$.
		\item $sol_2 > \max P_1$: the throughput decreases with $P_1 \Rightarrow P_{1}^*{= \min P_1}$.
		\item $sol_2 \in [\min P_1,\max P_1]$: as shown in the variation table of Fig. \ref{side2variationtable}, the throughput is decreasing between $\min P_1$ and $sol_2$, and increasing between $sol_2$ and $\max P_1$. Therefore, we obtain $P_1^* = \arg\max [F({\min P_1}), F({\max P_1})]$.
		\begin{figure}[H]
			\centering
			\scalebox{0.7}{
				\begin{tikzpicture}
				\tkzTabInit[lgt = 1, espcl=2.5]{$P_1$ / 1 , $\frac{\partial F}{P_1}$ / 1, $F$ / 1.5}{ $\min P_1$,  $sol_2$,$\max P_1$}
				\tkzTabLine{,-, z,+ }
				\tkzTabVar{+/ $F(\min P_1)$, -/ $F(sol_2)$, +/ $F(\max P_{1})$}
				\end{tikzpicture}
			}
			\caption{$R_{D2D}$ variation table when $sol_2 \in [\min P_1, \max P_1]$}
			\label{side2variationtable}
		\end{figure}
	\end{itemize}
	Finally, no matter if $\Delta>0 $ or $<0$, it is sufficient to test which of $\min P_1$ or $\max P_1$ delivers the best throughput and then select the corresponding segment endpoint.
	\subsection{Optimal Throughput point over the Side $S_U$}\label{app:S_u}
	Consider the sign of the sign of the polynomial's discriminant $\Delta = B^2 - 4AC$.
	If $\Delta < 0$: $\sign \partial F/ \partial P_1 = {\sign (h_d^2 - \eta_1\eta_2)}$.
	\begin{itemize}
		\item If $h_d^2 > \eta_1\eta_2, F$ is increasing with $P_1 \Rightarrow$ Set $P_1^{*}$ to $\max P_1$
		\item If $h_d^2 < \eta_1\eta_2, F$ is decreasing with $P_1 \Rightarrow$ Set $P_1^{*}$ to $\min P_1$
	\end{itemize}
	However, if $\Delta > 0$, then we have the two solutions 
	$sol_1 = (-B-\sqrt{\Delta})/2A$ and $sol_2 = (-B+\sqrt{\Delta})/2A$, with the  variation tables (Figs. \ref{fig:VarT_1} and \ref{fig:VarT_2}) depending on the sign of $h_d^2 - \eta_1\eta_2$. 
	\begin{itemize}
		\item If $h_d^2 > \eta_1\eta_2 \Rightarrow sol_1 < sol_2$ and $sol_1 < 0$. But not much can be said about the sign of $sol_2$ and how it compares to $\min P_1$ and $\max P_1$.
		\begin{figure}[H]\centering
			\scalebox{0.7}{
				\begin{tikzpicture}
				\tkzTabInit[lgt = 1, espcl=2.5]{$P_1$ / 1 ,  $\frac{\partial F}{\partial P_1}$ /1, $F$ /1.5}
				{$-\infty$, $sol_1$, $sol_2$, $+\infty$}
				\tkzTabLine{, +, z, -, z, +, }
				\tkzTabVar{-/ $-\infty$, +/ $F(sol_1)$, -/ $F(sol_2)$, +/ $+\infty$}
				\end{tikzpicture}
			}
			\caption{Variation table for $h_d^2 > \eta_1\eta_2$}\label{fig:VarT_1}
		\end{figure}
		However, we note that the right side of the variation table (where $P_1 > sol_1$) is similar to the variation table in Fig. \ref{side2variationtable}. Therefore, we conclude that:
		$$ P_1^{*} = \arg\max [ F(\min P_1), F(\max {P_1})].$$
		\commentaire{
			Based on the table of variation above we have this reasoning:
			\begin{itemize}
				\item If $sol_2<0$, Set $P_1$ to $\max P_1$
				\item If $sol_2>0$, then check
				\begin{itemize}
					\item  If $\min P_1 < sol_2 < \max P_1$, then set $P_{1}^* = \arg\max [L(\min P_1), L(\max P_1)]$
					\item If $sol_2 < \min P_1$, then set $P_1$ to ${\max P_1}$
					\item If $sol_2 > \max P_1$, then set $P_1$ to ${\min P_1}$.
				\end{itemize}
		\end{itemize}}
		\item if $h_d^2 < \eta_1\eta_2 \Rightarrow sol_2 < sol_1$ and $sol_1 > 0$.  
		\begin{figure}[H]\centering
			\scalebox{0.7}{
				\begin{tikzpicture}
				\tkzTabInit[lgt = 1, espcl=2.5]{$x$ / 1 ,  $\frac{\partial F}{\partial P_1}$ /1, $F$ /1.5}
				{$-\infty$, $sol_2$, $sol_1$, $+\infty$}
				\tkzTabLine{, -, z, +, z, -, }
				\tkzTabVar{+/ $\infty$, -/ $F(sol_2)$, +/ $F(sol_1)$, -/ $-\infty$}
				\end{tikzpicture}
			}
			\caption{Variation table for $h_d^2 < \eta_1\eta_2$}\label{fig:VarT_2}
		\end{figure}
		Since $sol_1$ is a local maximum, $F(sol_1) > F(P_1), \forall P_1 > sol_2$. Then, the only values of $P_1$ which might give a better throughput than $sol_1$ are those at the left of $sol_2$. We can distinguish the following three cases:
		\begin{itemize}
			\item if $\max P_1 < sol_1$, set $P_1^{*}$ to $\max P_1$. 
			\item if $ sol_1 < \min P_1$, set $P_1^{*}$ to $\min P_1$. 
			\item if $sol_1 \in [\min P_1, \max P_1]$, then:
			\begin{itemize}
				\item if $\min P_1 > sol_2$, set $P_{1}^{*}$ to $sol_1$.
				\item if $\min P_1 < sol_2$, set\\ $P_{1}^{*} = \arg\max [ F(\min P_1), F(sol_1)].$
			\end{itemize}
		\end{itemize}	
	\end{itemize}
	To sum up, in the optimization over the intersection segment of $\mathcal{PL}_2$ with $S_U$, all the possible channel conditions lead at some point to choosing $P_{1}^{*}$ from the values $\min P_1$, $\max P_1$, and $sol_1$ (when it is included in the interval $\mathbb{U}$) according to the one delivering the highest throughput. 
	\subsection{Endpoint Coordinates}\label{app:EndPoints}
	The coordinates of the segment endpoints $k_1,k_2,j_2,x_1$, $x_2,x_U,s_1,s_2,s_U,v_1,v_2,v_3,v_4,v_5,g_1,g_2,g_3,g_4,g_5$ are given below. Note that $x_i$ and $xl_i$ (resp. $s_i$ and $sl_i$)  have the same expressions with the difference that $xl_i$ (resp. $sl_i$) is not defined outside of $S_i$. Moreover,  $x_i$ and $s_i$ have strictly positive coordinates since $h_{d_1,u}h_{b,d_2} - h_{b,u}h_d > 0$ from eq. \eqref{necessary2}, and $h_{b,d_1}h_{d_1,u} - \eta_1 h_{b,u}>0$  from eq \eqref{necessary3}.
	\begin{table}[H]
		\begin{tabular}{l}
			$k_1=(P_{1,M},(P_{u,m}h_{bu} + P_{1,M}h_{b,d_1})/h_{b,d_2},P_{u,m})$,  
			\\
			$k_2=(	(P_{2,M}h_{b,d_2} - P_{u,m}h_{b,u})/h_{b,d_1},P_{2,M},P_{u,m})$,
			\\
			$j_2=(	P_{u,m} h_{b,u}/h_{b,d_1},P_{2,M},P_{u,m})$,
			\\
			$x_1=(1,
			\dfrac{h_{b,u} \eta_1 + h_{d_1,u} h_{b,d_1}}{h_{d_1,u} h_{b,d_2} - h_{b,u} h_d},
			\dfrac{h_{b,d_2} \eta_1 + h_{d} h_{b,d_1}}{h_{d_1,u} h_{b,d_2} - h_{b,u} h_d}  ) P_{1,M}$,
			\\
			$x_2= (\dfrac{h_{d_1,u} h_{b,d_2} - h_{b,u} h_d}{h_{b,u} \eta_1 + h_{d_1,u} h_{b,d_1}},1,\dfrac{h_{b,d_2} \eta_1 + h_{d} h_{b,d_1}}{h_{b,u} \eta_1 + h_{d_1,u} h_{b,d_1}})P_{2,M}$,    
			\\
			$s_1= (1,\dfrac{h_{b,d_1} h_{d_1,u} - \eta_1 h_{b,u}}{h_{b,u} h_d},\dfrac{h_{b,d_1}}{h_{b,u}})P_{1,M}$,
			\\
			$s_2= (\dfrac{h_{b,u} h_d}{h_{b,d_1} h_{d_1,u} - \eta_1 h_{b,u}}, 
			1,\dfrac{h_{b,d_1} h_d}{h_{b,d_1} h_{d_1,u} - \eta_1 h_{b,u}})P_{2,M}$,
			\\
			$s_u = (\dfrac{h_{b,u}}{h_{b,d_1}},\dfrac{h_{b,d_1} h_{d_1,u} - \eta_1 h_{b,u}}{h_{b,d_1} h_d},1) P_{u,M},$
			\\
			$x_u = (\dfrac{h_{d_1,u} h_{b,d_2} - h_{b,u} h_d}{h_{b,d_2} \eta_1 + h_{d} h_{b,d_1}},
			\dfrac{h_{b,u} \eta_1 + h_{d_1,u} h_{b,d_1}}{h_{b,d_2} \eta_1 + h_{d} h_{b,d_1}},
			1)P_{u,M}.$
		\end{tabular}
	\end{table}
	The $w_i$ family is obtained by combining $v_i$ and $g_i$.
	\begin{table}[H]
		\begin{tabular}{l}
			$v_1=(P_{1,M},  (P_{u,m} h_{d_1,u} - P_{1,M}\eta_1)\frac{1}{h_{d}}, P_{u,m})$,  \\
			$v_2=((P_{u,m}h_{d_1,u} - P_{2,M}h_{d})\frac{1}{\eta_1},P_{2,M},P_{u,m})$,      \\
			$v_3=(P_{1,M}, P_{2,M}, (P_{1,M} \eta_1 + P_{2,M} h_{d}) \frac{1}{h_{d_1,u}})$, \\
			$v_4=(P_{1,M},  (P_{u,M} h_{d_1,u} - P_{1,M}\eta_1)\frac{1}{h_{d}}, P_{u,M})$,  \\
			$v_5= ((P_{u,M}h_{d_1,u} - P_{2,M}h_{d})\frac{1}{\eta_1},P_{2,M},P_{u,M}) $,    \\
			$g_1=(P_{1,M},  (P_{u,m} h_{d_2,u} - P_{1,M}h_{d})\frac{1}{\eta_2}, P_{u,m}) $, \\
			$g_2=((P_{u,m}h_{d_2,u} - P_{2,M}\eta_2)\frac{1}{h_{d}},P_{2,M},P_{u,m})$,      \\
			$g_3=(P_{1,M}, P_{2,M}, (P_{1,M} h_{d} + P_{2,M} \eta_2) \frac{1}{h_{d_2,u}})$, \\
			$g_4=(P_{1,M},  (P_{u,M} h_{d_2,u} - P_{1,M}h_{d})\frac{1}{\eta_2}, P_{u,M})$,  \\
			$g_5=((P_{u,M}h_{d_2,u} - P_{2,M}\eta_2)\frac{1}{h_{d}},P_{2,M},P_{u,M}).$
		\end{tabular}
	\end{table}

\end{document}